\documentclass[wcp]{jmlr}

\usepackage{longtable}

\usepackage{booktabs}
\usepackage{graphicx}
\usepackage{orcidlink}
\usepackage{multirow}
\usepackage{amssymb}
\usepackage{makecell}
\usepackage{mathtools}
\usepackage{subcaption}
\usepackage{array}

\makeatletter
\let\Ginclude@graphics\@org@Ginclude@graphics 
\makeatother

\jmlrvolume{}
\jmlryear{2024}
\jmlrworkshop{ACML 2024}

\title[Efficient Federated Unlearning]{Efficient Federated Unlearning under \\ Plausible Deniability}
\author{\Name{Ayush K. Varshney} \Email{ayushkv@cs.umu.se} \\
        \Name{Vicen\c{c} Torra} \Email{vtorra@cs.umu.se}\\
        \addr Department of computing science, Umeå University, Sweden}

\editors{Vu Nguyen and Hsuan-Tien Lin}

\begin{document}

\maketitle

\begin{abstract}
Privacy regulations like the GDPR in Europe and the CCPA in the US allow users the right to remove their data from machine learning (ML) applications. Machine unlearning addresses this by modifying the ML parameters in order to forget the influence of a specific data point on its weights. Recent literature has highlighted that the contribution from data point(s) can be forged with some other data points in the dataset with probability close to one. This allows a server to falsely claim unlearning without actually modifying the model's parameters. However, in distributed paradigms such as federated learning (FL), where the server lacks access to the dataset and the number of clients are limited, claiming unlearning in such cases becomes a challenge. An honest server must modify the model parameters in order to unlearn. This paper introduces an efficient way to achieve machine unlearning in FL, i.e., federated unlearning, by employing a privacy model which allows the FL server to plausibly deny the client's participation in the training up to a certain extent. Specifically, we demonstrate that the server can generate a \textit{Proof-of-Deniability}, where each aggregated update can be associated with at least $x$ (the plausible deniability parameter) client updates. This enables the server to plausibly deny a client's participation. However, in the event of frequent unlearning requests, the server is required to adopt an unlearning strategy and, accordingly, update its model parameters. We also perturb the client updates in a cluster in order to avoid inference from an honest but curious server. We show that the global model satisfies $(\epsilon, \delta)$-differential privacy after $T$ number of communication rounds. The proposed methodology has been evaluated on multiple datasets in different privacy settings. The experimental results show that our framework achieves comparable utility while providing a significant reduction in terms of memory ($\approx$ 30 times), as well as retraining time (1.6-500769 times). The source code for the paper is available \href{https://github.com/Ayush-Umu/Federated-Unlearning-under-Plausible-Deniability}{here}.
\end{abstract}

\section{Introduction}
The recent surge in artificial intelligence (AI) and machine learning (ML) has significantly impacted various sectors, including healthcare, finance, transportation, and day-to-day life in general. ML analyzes data collected from individual subjects in order to learn from them. The information from this data is encoded within the weights of ML models.  Recognizing the potential for misuse of both the data and the information these models encode, AI regulatory frameworks, such as the General Data Protection Regulation (GDPR) and the California Consumer Privacy Act (CCPA), have been established. These regulations allow users the right to have their data removed, protecting their privacy.

The easiest approach to forget the user(s) in centralized ML (i.e., Machine Unlearning) is to delete the user from the database and retrain the model from scratch. This approach is costly and time-consuming, making it impractical. A more careful approach should aim to remove user data without incurring the retraining cost from scratch. The objective is to produce a model identical to what one would achieve by training on the dataset after excluding the data point meant to be forgotten. This requirement is quite strict; however, ~\cite{ginart2019making} introduced a more flexible concept of unlearning in a centralized ML environment, called approximate unlearning. In approximate unlearning, the model owner updates the existing model parameter slightly (e.g. using gradient ascent~\cite{halimi2022federated}, knowledge distillation~\cite{wu2022federated}) to obtain parameters similar to a model naively retrained on a dataset which does not have the information from the user to be unlearned. ~\cite{thudi2022necessity} argue that this definition of unlearning is ill-defined. The authors introduce the concept of \textit{Proof-of-Learning} (PoL), in which a minibatch can be forged (i.e., generating similar gradient updates) by a set of minibatches which do not have the data point(s) to be unlearned. Furthermore,~\cite{kong2022forgeability} show the connection between membership inference attack and machine unlearning by claiming that the data owner can leverage PoL to provide \textit{Proof-of-Repudiation} (PoR) which repudiates the claim of MIA. In MIA, an attacker aims to identify whether a data point participated in training the ML model or not. Kong et al. considers a scenario in which the adversary carries out MIA and correctly predicts whether \textit{x} (data point to be unlearned) participated in training. The PoR enables the data owner to repudiate the claims of MIA by providing a set of forging minibatches, which results in similar gradient updates. 


Both approaches consider a centralized ML environment where the data owner has complete access to the dataset in order to create forged minibatches. However, in a distributed paradigm such as federated learning (FL), which offers joint training of the global model using multiple clients without sharing their data, the assumption of data availability does not hold. In FL, clients share the model parameters with the central server and the server aggregates them to train a global model. This process continues for few communication rounds until the desired results are obtained. In FL, either a user can ask the central server to be unlearned or a user can ask to remove the influence of some of their data points. Here, the server is responsible for learning as well as unlearning without access to the dataset. In the absence of data, PoL and PoR can not be used to create forged minibatches (\cite{thudi2022necessity},~\cite{kong2022forgeability}) to avoid unlearning in FL. 

The first unlearning approach in FL was proposed in~\cite{liu2021federaser} where the server stores the model updates from each client. When a client or a group of clients request for unlearning, the server retrains the model using the stored updates and remaining clients. This approach requires a huge amount of storage at the server side, and in addition, has high cost of retraining. Similarly, ~\cite{wu2022federated} also store the client updates on server, and subtract the historical updates of the targeted clients from the global model. Then, the knowledge distillation is used with synthetic data to train the skewed model. The accuracy in such an unlearning method is negatively affected by the degree of non-iid data. ~\cite{liu2022right} propose approximate unlearning using first order Taylor expansion which requires participation from all the clients. ~\cite{halimi2022federated} propose retraining the target client with gradient ascent to maximizing the loss on its local data before deletion. The idea that gradient ascent can lead to unlearning or removing the influence of data points seems bogus. ~\cite{wang2023federated} highlight the potential privacy risks in federated unlearning and recommend that privacy-preserving mechanisms should be incorporated while unlearning. Most of the work in the literature on federated unlearning is computationally expensive and requires a high amount of storage on the server. The retraining cost becomes increasingly impractical when there are large number of clients. 

In order to overcome these drawbacks, we focus on generating models which can be generated by multiple sets of clients in each communication round i.e., models which recurs from multiple sets of clients. This allows the central server to avoid employing unlearning mechanisms for every unlearning request. The generation of recurring models can be achieved through integral privacy. ~\cite{varshney2023integrally} show a methodology for generating integrally private deep neural networks. The generated private models have comparable utility with non-private models, however, the recurrence of the models is probabilistic. ~\cite{varshneyk} show that, under similar training environment and a large batch size, the model trained from clients having data sampled from a set of distributions will likely have their gradient updates separated by only a small distance (say $\Delta$), with high probability. This indicates that for each communication round, there exist multiple sets which generate similar gradient updates with mean-sampling optimizers such as stochastic gradient descent (SGD), Adam, etc. In simple words, there are no one-to-one mapping(s) between the gradient and the clients. 

In each communication round, the approach described in \cite{varshneyk} clusters the clients and randomly chooses a representative from each cluster for the global model aggregation. In such a scenario, the server can plausibly deny the participation of the targeted client ($c^*$) in training, provided that there are at least $x-1$ (where $x$ is the plausible deniability parameter) different clients within the cluster that have similar gradient updates. In this paper, we demonstrate that by adopting integrally private federated averaging~\cite{varshneyk}, the server can produce a \textit{Proof-of-Deniability} (PoD), whereby the server can provide a log of training which does not contain the target client in the training of the current global model. This approach benefits with large number of clients as a weight can be mapped to many clients. Historically, if the number of clients generating similar model updates are less than $x$ in any cluster, the server must employ the unlearning mechanism. Furthermore, we introduce a client-level differentially private mechanism to select a cluster representative in order to protect client's identity in each cluster from honest but curious server along with its privacy analysis. 

In summary, we make the following contributions.

\begin{enumerate}
    \item A novel federated unlearning framework in which the server can provide the PoD to deny a clients' participation in the training.
    \item A client-level differentially private mechanism to protect the identity of the participating client during aggregation from honest but curious server.
    \item A theoretical analysis showing that the global model satisfies $(\epsilon, \delta)$-differential privacy for $0< \epsilon < 8\log(1/\delta)$ and $\delta>0$ if and only if $\sigma^2 \geq \frac{T(1+8\log(1/\delta))}{7\epsilon^2}$.
    \item Empirical results proving the computational efficiency ($1.6-500769$ times) along with the significant improvement in the memory storage at the server ($\approx 30$ times) of our framework when retraining is used as the unlearning mechanism.
\end{enumerate}

\section{Background}

In this section, we provide the details of the background knowledge needed in this work.

\subsection{Integral Privacy} \label{Integral privacy}

The integral privacy model by \cite{torra2020explaining} was initially proposed as a defense against the model comparison attack and membership inference attack. Simply, integrally private models are the models which recur multiple times (number of recurrence is application dependent), where models are trained on different subsamples which do not share records among them. The condition to not share records among samples is required to avoid any inference using intersection analysis. But with the large number of weights in DNNs, generating exactly the same model is very computationally expensive. In \cite{varshney2023integrally}, a relaxed notion called $\Delta$-Integral privacy ($\Delta$-IP) was proposed where models at most $\Delta$ distant apart were considered equivalent. Formally, $\Delta$-IP can be defined as follows.


\textbf{$\Delta$-Integral Privacy} Let $\mathcal{D}$ be the population, $S^* \subset \mathcal{D}$ be the background knowledge, and $M \subset \mathcal{M}$ be the model generated by an algorithm $A$ on an unknown dataset $X \subset \mathcal{D}$. Then, let $Gen^*(M, S, \Delta)$ represent the set of all generators consistent with background knowledge but not including $S^*$ and model $M$ or models at most $\Delta$ distant. Then, $k$-anonymity $\Delta$-IP holds when $Gen^*(M, S, \Delta)$ has at least $k$-elements and, 
\begin{equation} \label{e1.1}
\bigcap_{S \in Gen^*(G, S^*, \Delta)} S = \emptyset.
\end{equation} 

\subsection{Federated Learning} \label{sec:federated learning}
\begin{figure}
    \centering
    \includegraphics[width=0.6\textwidth]{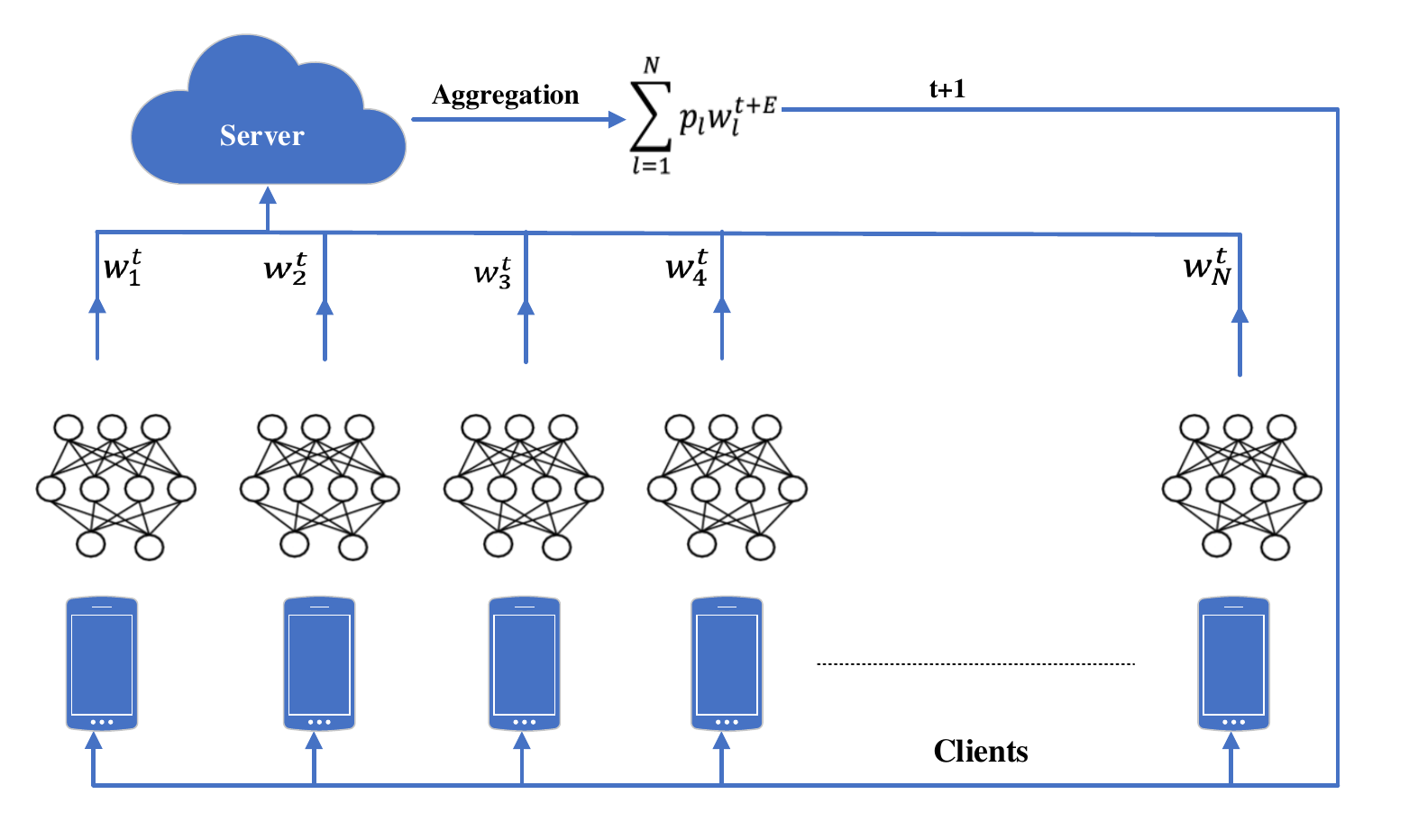}
    \caption{Federated learning framework using fedAvg algorithm.}
    \label{fig:fedAvg}
\end{figure}

In federated learning (see Fig.~\ref{fig:fedAvg}), a central server initializes the global model. At each communication round, the server communicates the global model to the participating clients. The clients train the global model on their data for few epochs and communicate the updated model to the central server. The central server aggregates the updated models from participating clients and this process continues for a given number of communication rounds~\cite{mcmahan2017communication}. In full-device participation, all the clients in the network participate to train the global model in each communication round, on the other hand in partial-device participation, few randomly chosen clients participate to train the global model. The typical federated optimization for the server looks like:

\begin{equation} 
    \min_{w} \left \{  F(w) \triangleq \sum_{l=1}^{N} p_l F_l(w) \right \} \label{FLobj}
\end{equation}
where $N$ is the number of clients, $p_l \; (p_l \geq 0 \And \sum_{l=1}^N p_l = 1)$ is the weight of $l^{th}$ client and $F_l(w)$ is the local objective function to minimize the loss on the local data with weight $w$. For a user-specific loss function (say $l()$), suppose the $l^{th}$ device has $n_l$ number of training instances ($(x_1, y_1), (x_2, y_2)....,(x_{n_l}, y_{n_l})$). Then, the local objective function $F_l(w)$ can be defined as:
\begin{equation}
   \min_w F_l(w) \triangleq \frac{1}{n_l}\sum_{i=1}^{n_l} l(w; x_i, y_i)
\end{equation}

In Fig.~\ref{fig:fedAvg}, $t$ is the communication round, $w_l^t$ is the weight of the client $l$, $w^t$ is the global model at $t^{th}$ communication round, and $\xi_l^{t+e}$ is a sample uniformly chosen from $l^{th}$ client's local data.

\subsection{Membership Inference Attack} \label{secrity game}

Membership Inference Attacks (MIA) aim to predict whether a data point participated in training a given machine learning model. The machine unlearning literature widely uses MIA to audit whether the ML model has unlearnt the target client or not. 

There have been several attempts in the literature~\cite{carlini2022membership, jayaraman2020revisiting, kong2022forgeability} which formalize MIA as a security game. The game evaluates privacy leakage, and it is played between a challenger ($\mathrm{Ch}$) and an adversary ($\mathcal{A}$). The challenger (dataset owner) challenges the adversary with background knowledge ($S^*$), to predict whether a data sample participated in the training or not. The positive outcome of the game determines the success of the attack. The game in ~\cite{jayaraman2020revisiting} $\mathcal{SG}_{MI}(.)$ is played as:

\begin{enumerate}
    \item The data owner acting as challenger, $\mathrm{Ch}$, samples a training dataset ($\mathcal{D}$) from the original dataset ($\mathbb{D}$) and trains a machine learning model $\mathcal{M}$ with it.
    \item The challenger $\mathrm{Ch}$ randomly selects $b$ from $\{0,1\}$. If $b=1$, then $\mathrm{Ch}$ samples a data point ($x, y$) from $\mathcal{D}$, otherwise $\mathrm{Ch}$ samples $(x, y)$ from $\mathbb{D}\setminus\mathcal{D}$.
    \item $\mathrm{Ch}$ sends $(x, y)$ to the adversary $\mathcal{A}$.
    \item The adversary evaluates $\mathcal{A}((x,y), S^*, \mathcal{M})$, i.e., decides whether the sample $(x,y)$ participated in the training or not.
    \item Return 1 if $\mathcal{A}((x,y), S^*, \mathcal{M}) = b$, 0 otherwise. 
\end{enumerate}

The security game can be leveraged to audit the unlearning, it can be modified to return 1 if the sample $(x,y)$ is part of the training set, otherwise return 0. 

\subsection{Forgeability and Proof-of-Learning}

Forgeability introduced by~\cite{thudi2022necessity} has been given in the context of datasets, i.e., two datasets are called forgeable if they produce similar model updates which are at most $\epsilon$ ($\epsilon << 1$) distance apart. The small deviation is allowed as a result of some per-step error due to optimization in the mean samplers (e.g., SGD, Adam). This is specifically useful in the context of machine unlearning where the data owner stores training logs. Training logs consist of a sequence of data points from the dataset $\mathbb{D}$ and their gradient updates from the mean sampler as check points. This acts as \textit{Proof-of-Learning} for the model $\mathcal{M}$. When an unlearning request comes, say for the data point $x$, the data owner forges the minibatches containing the data point $x$ and produces a \textit{Proof-of-Repudiation} (\cite{kong2022forgeability}), claiming the absence of $x$ during training. Formally, forgeability is defined below.

\textbf{Forgeability:} Two datasets $\mathcal{D}, \mathcal{D}'$ are said to forgeable if for the model $\mathcal{M}$ we have,
\begin{align}
    \forall x_i \in \mathcal{D}, \exists \; \Bar{x}_i \in \mathcal{D}',\text{ such that,} \\ 
    ||g(\mathcal{M}, x_i) - g(M, \Bar{x}_i)||_2 \leq \epsilon
\end{align}
here, $g$ is the model update rule. The idea behind this definition is that the parameter update due to any minibatch in $\mathcal{D}$ can be mapped to at least one minibatch in $\mathcal{D}'$. Now, in order to repudiate the membership claim from the MIA security game $\mathcal{A}((x,y), S^*, \mathcal{M})$ defined in Section~\ref{secrity game}, the data owner (or the challenger) finds functionally equivalent models. 

\textbf{Functional Equivalence}~\cite{kong2022forgeability}: Two models {$\mathcal{M}, \mathcal{M}'$} are said to be functionally equivalent with respect to the adversary $\mathcal{A}$ for a given dataset $\mathcal{D}$ if and only if,
\begin{align}
    \forall (x, y) \in \mathcal{D}, \; \mathcal{A}((x,y), S^*, \mathcal{M}) = \mathcal{A}((x,y), S^*, \mathcal{M}')
\end{align}
Intuitively, this means that with respect to MIA security game, an adversary predicts same predictions for functionally equivalent models on all the data points in $\mathcal{D}$. To allow some small error step, the following conjecture has been given.  

\textbf{Conjecture 1}~\cite{kong2022forgeability}: Two models, $\mathcal{M}, \mathcal{M}'$ are functionally equivalent with respect to MIA iff, $||\mathcal{M} \ominus \mathcal{M}'|| \leq \epsilon$ and $\epsilon$ is a small value. 

\noindent Here, this conjecture allows the data owner to repudiate the claims of MIA security game and hence plausibly deny the participation of the targeted data point(s).  

\subsection{Differential Privacy} \label{sec: diff priv}

Differential privacy (DP) is a widely accepted privacy framework. The classical definition of $(\epsilon, \delta)$-DP is given below.

\textbf{$(\epsilon, \delta)$-Differential privacy:} For two neighbouring dataset $D_1, D_2$, privacy parameter $\epsilon > 0$ and $0 \leq \delta < 1$, a function $f_r$ for query $r$ is considered $(\epsilon, \delta)$-differentially private iff,
\begin{align}
    Pr[f_r(D_1)\in S] \leq e^\epsilon Pr[f_r(D_2) \in S] + \delta
\end{align}
where $S \subseteq Range(f_r)$. The composition of privacy budget in DP over multiple iterations is not straightforward. R\'enye differential privacy (RDP) was proposed to overcome this drawback. 

\textbf{R\'enye differential privacy:} For two neighbouring dataset $D_1, D_2$, privacy parameter $\rho \geq 0$ and $\alpha > 0$, then a function $f_r$ over query $r$ satisfies $(\alpha, \rho)$-RDP if the $\alpha$-divergence between them satisfies:
\begin{align}
    D_{\alpha}[F_r(D_1)||f_r(D_2)] = \frac{1}{\alpha-1} \log \mathbb{E}\left[ \left(\frac{f_r(D_1)}{f_r(D_2)}\right)^{\alpha}\right] \leq \rho(\alpha)
\end{align}
RDP is a relaxed version of DP which provides tighter composition bound. The $(\alpha, \rho(\alpha))$-RDP can be converted into $(\epsilon, \delta)$-DP using the following lemma.

\textbf{Lemma 1.}~\cite{mironov2017renyi} If the function satisfies $(\alpha, \rho(\alpha))$-RDP, then it also satisfies $\left( \rho(\alpha) + \frac{\log(1/\delta)}{\alpha-1}, \delta \right)$-DP $\forall \; 0< \delta < 1$.

We will also use the following definitions and lemmas to derive the privacy analysis of our methodology.

\textbf{$l_2$-sensitivity:}  For the function $f_r$, the $l_2$ sensitivity  $\psi(f_r)$ is defined as: $\psi(f_r) = \max||f_r(D_1) - f_r(D_2)||_2$

\textbf{Lemma 2.}~\cite{mironov2017renyi} Let $f_r$ be the query function with $l_2$ sensitivity $\psi(f_r)$. The Gaussian perturbation given by: $GM = f_r(D) + N (0, \sigma^2\psi(f_r)^2\mathbb{I})$ satisfies $(\alpha, \frac{\alpha}{2\sigma^2})$-RDP.

\textbf{Lemma 3.}~\cite{mironov2017renyi} Let $f_r^1, f_r^2$ represent two query functions on a dataset $D$ satisfying $(\alpha, \rho_1(\alpha))$-RDP and $(\alpha, \rho_2(\alpha))$-RDP. Then their composition $f_r^1 \circ f_r^2$ satisfies $(\alpha, \rho_1(\alpha) + \rho_2(\alpha))$-RDP. 

\subsection{Plausible Deniability}

An algorithm satisfies plausible deniability~\cite{bindschaedler2017plausible} if a set of records can independently generate a given output with a certain probability bound. This results in input indistinguishability for an intruder with background information who is looking to infer if a particular record is significantly more responsible for the output. ~\cite{bindschaedler2017plausible} defined plausible deniability as follows.

\textbf{Plausible Deniability:} Let $\mathcal{D}$ be a dataset having at least $x$ number of records, then for a given output $y$ by model $\mathcal{M} \text{ i.e. } y = \mathcal{M}(d_1) \text{ for } d_1 \in \mathcal{D}$, we say that model $\mathcal{M}$ satisfies $(x, \gamma)$ plausible deniability if there exists at least $x-1$ distinct records ($d_2, d_3, ...,d_x \in \mathcal{D} \setminus \{d_1\}$) such that:

\begin{equation}
    \gamma^{-1} \leq \frac{Pr[\mathcal{M}(d_i) = y]}{Pr[\mathcal{M}(d_j) = y]} \leq \gamma \label{plausible}
\end{equation}


\section{Proposed Work}

In this section, we provide the details of the proposed plausibly deniable unlearning framework for FL. The existing work in the literature of forging~\cite{thudi2022necessity, kong2022forgeability} considers the availability of datasets and assumes freedom over $\mathbb{D}$ to sample minibatches indefinitely. These assumptions are not valid for federated learning, i.e., the central server does not has access to the dataset and the number of participating clients are limited. Also, both of the approaches in the literature (PoL, and PoR) consider unlearning a single sample. In this work, we consider the request for unlearning to be a continuous phenomenon and a client can request for unlearning at any communication round. 

Consider a typical FL scenario with limited number of clients, and frequent unlearning requests. In such cases, employing unlearning mechanisms such as retraining, or any approximate unlearning mechanism frequently can be computationally costly and requires huge storage at the central server. This necessitates the exploration of plausibly deniable unlearning solutions in FL. Since the number of clients are limited, the plausible deniable solutions are effective up to a certain degree. And in case of frequent unlearning requests, the central server will eventually need to employ an unlearning mechanism. Inspired by PoL, and PoD, we propose the concept of \textit{Proof-of-Deniability} (PoD) for plausibly denying a client participation in federated learning. 

Next, we delve into the MIA security game (refer to Section~\ref{secrity game}) to explore how it can be leveraged to audit the unlearning of a client in federated learning. In each communication round in FL, the server samples $N$ clients weights ($C=\{ c_1, c_2,...,c_N \}$) from the the set of all client weights $\mathbb{C} = \{c_1, c_2, ..., c_S\}$, $S$ be the total number of clients participating in the FL, to train the global model (see Section~\ref{sec:federated learning}) which is then communicated to all the clients. This process continues for $T$ rounds. In order to audit unlearning using the MIA security game in the communication round $t$, the challenger $\mathrm{Ch}$ (unlearner) receives a client weight (say $c^*$) and employs an unlearning mechanism to remove the influence of $c^*$. The challenger communicates the updated model (say $\mathcal{G}$) to the adversary. The adversary (or auditor) $\mathcal{A}$ with background knowledge ($S^*$) tries to find whether $c^*$ participated in the training of $\mathcal{G}$ or not. The game returns, whether the unlearner removed the influence of $c^*$ or not. Formally, the security game for FL $(\mathcal{SG}_{FL}())$ is defined as follows:
\begin{enumerate}
    \item The challenger $\mathrm{Ch}$ receives a client weight $c^* \in \mathbb{C}$. 
    \item The challenger removes the contribution of $c^*$ on $\mathcal{G}$ with some probability $(\leq 1).$
    \item $\mathrm{Ch}$ sends $c^*$ along with the updated global model $\mathcal{G}'$ to adversary $\mathcal{A}$.
    \item The adversary evaluates $\mathcal{A}(c^*, S^*, \mathcal{G}') \rightarrow \{0,1\}$~\cite{suri2022subject}, i.e., whether the client $c^*$ participated in the training of $\mathcal{G}'$.
    \item Return 1 if $\mathcal{A}(c^*, S^*, \mathcal{G}) = 1$, 0 otherwise.
\end{enumerate}

\begin{algorithm}
\DontPrintSemicolon  
\caption{Perturbed $k$-Anonymous Integrally Private Federated Averaging}
\label{Algo IPfedAvg}

\SetKwInput{KwInput}{Input}                
\SetKwInput{KwOutput}{Output}              
\rule{\linewidth}{0.1pt}

\textbf{Server side:}\;
Initialize global model $w_0$\;
\For{$t = 1,2,\dots, \lfloor \frac{T}{E} \rfloor$} {  
    Broadcast $w_t^g$ to all the clients\;
    \For{each client $l = 1, 2, \dots, N$}{
        $w_{t+1}^l \leftarrow \textbf{ClientUpdate}(w_t)$\;
    }
    Compute clusters $C = \{C_1, C_2, \dots, C_{\lfloor \frac{N}{k} \rfloor}\}$ based on some $\Delta$ parameter \\
    Perturb randomly chosen model:    $w_{t+1}^{'r_c} = w_{t+1}^{r_c} + N(0, \sigma^2\Delta^2\mathbb{I})$ \\
    $w_{t+1}^g = \sum_{c=1}^{|C|} p_cw_{t+1}^{'r_c}$ \tcp*{Aggregate perturbed models}
    Server stores the index of clients in each cluster and $w_{t+1}^g$\;
}
\rule{\linewidth}{0.1pt}
\textbf{ClientUpdate}$(w_t)${
    \KwInput{Initial weight $w_t$}
    \KwOutput{Updated weight $w$}
    Consider $w = w_t$ as initial weight\;
    \For{local epochs $e = 1, 2, \dots, E$}{
        $w \leftarrow w - \eta_t \nabla F_l(w, \xi_l^{t+e})$\;
    }
    \Return{$w$}\; \\
}
\rule{\linewidth}{0.1pt}
\end{algorithm}

\subsection{Proof-of-Deniability}

In this section, we propose a methodology in which the server can provide the \textit{Proof-of-Deniability} to refuse membership inference claim of $\mathcal{SG}_{FL}()$. In our methodology, in each communication round the central server clusters the clients' weights according to some distance measure. Then it randomly chooses a representative from each cluster, perturbs it based on the integral privacy parameter ($\Delta$) and then aggregates these perturbed representatives to generate the global model for the next communication round~\cite{varshneyk}. This approach is the perturbed variation of 'k-Anonymous Integrally Private Federated Averaging' (perturbed $k$-IPfedAvg). 

In perturbed $k$-IPfedAvg, when the server receives an unlearning request, the id of the target client(s) and its associated weight are removed from the clusters in each round. Server rollbacks and retraining is performed only when any cluster has less than the predefined number of clients ($x$ in plausible deniability). Plausible deniability in Eq. (\ref{plausible}) is given for records of a dataset. We now define plausible deniability for client participation where $x$ clients generate similar model updates. 

\textbf{Plausible Deniability for client participation:} Let $N$ be the number of clients, with model weights $W_t = \{w_t^1, w_t^2,..., w_t^N\}$, participating in training the global model at $t^{th}$ communication round. For a given model weight, say $w_t^i$, a server can $(x, \Delta)$ plausible deny the client participation if there exist a set of at least $x-1$ distinct client weights $(w_t^j)$ such that:
\begin{equation}
||w_t^i - w_t^j||_2 \leq \Delta
\end{equation}

\begin{figure}
    \centering
    \includegraphics[width=0.8\textwidth]{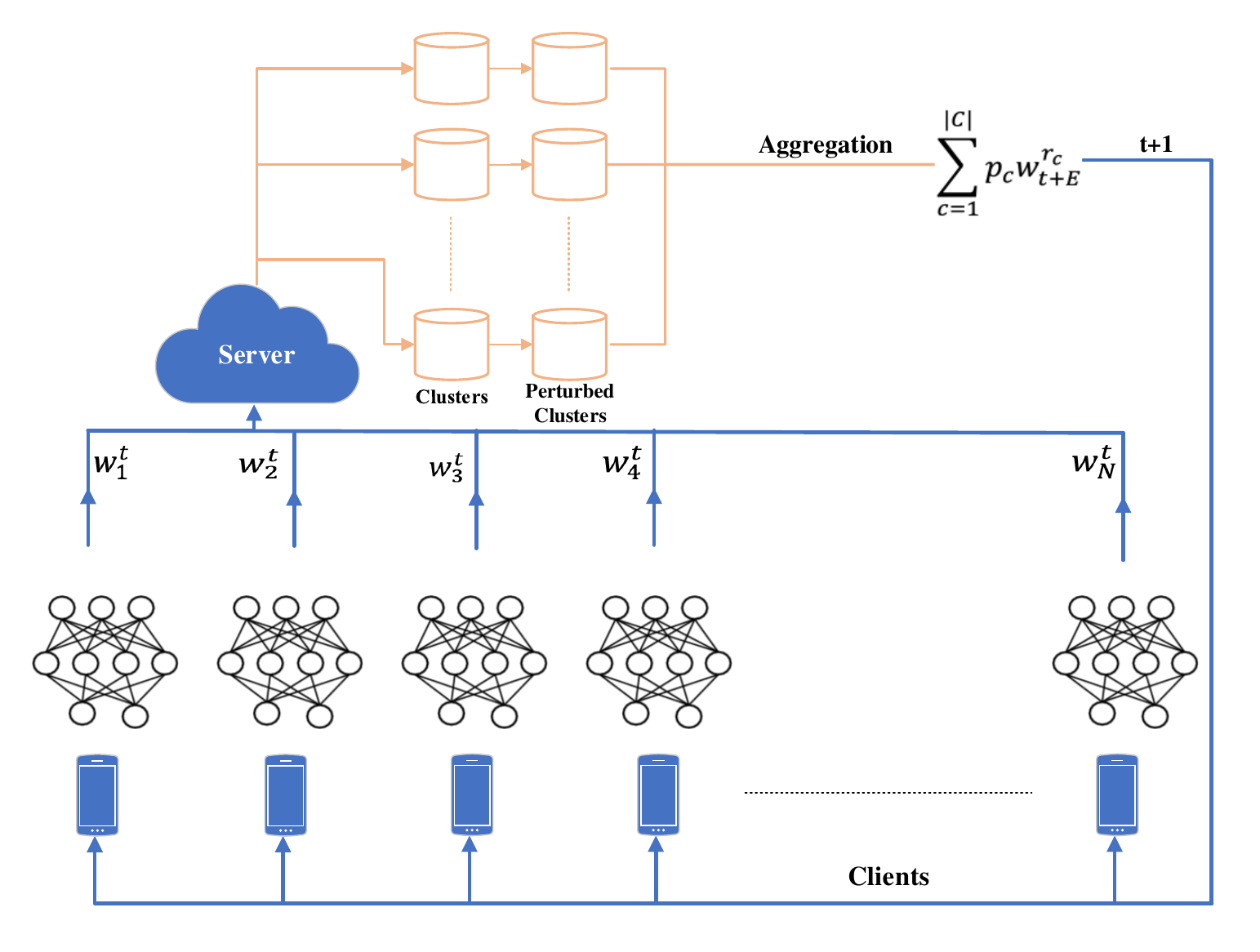}
    \caption{Efficient federated unlearning framework using $k$-IPfedAvg algorithm.}
    \label{fig:Efficient unlearning}
\end{figure}

Algorithm~\ref{Algo IPfedAvg} provides the formal algorithm for perturbed $k$-Anonymous integrally private privacy-preserving averaging. In the $t^{th}$ communication round, first the server broadcasts the current global model (say $w_t^g$) to all the clients. Then, the server clusters the model updates from the clients ($w_{t+1}^1, w_{t+1}^2, ..., w_{t+1}^N$) based on the predefined threshold (say $\Delta$). For a small $\Delta$, the model updates in each cluster will be very similar i.e. we can say that the models in each cluster are $\Delta$-integrally private~\cite{varshney2023integrally}. In $k$-IPfedAvg, the server randomly chooses one model from each cluster as their cluster representative and aggregate them to compute new global model as:
\begin{equation}
    w_{t+1}^g = \sum_{l=1}^{|C|} p_c w_{t+E}^{r_c}
\end{equation}
where $w_{t+E}^{r_c}$ is a randomly chosen model weight from each cluster and $p_c = \sum_{i \in C_c} p_i$. Here, an honest but curious server will have access to which client has been used as cluster representative. We further remove this drawback with client-level $(\epsilon, \delta)$-differential privacy~\cite{geyer2017differentially}. In order to fully avoid inference of any particular client i.e. to make model weights indistinguishable, the randomly chosen representative from each cluster is perturbed with noise. The server adds Laplacian or Gaussian noise based on the integral privacy parameter($\Delta$).


Then, the aggregated global model is computed as:
\begin{equation}
    w_{t+1}^g = \sum_{l=1}^{|C|} p_c w_{t+E}^{r'_c} = \sum_{l=1}^{|C|} p_c \left(  w_{t+E}^{r_c} + N(0,\sigma^2\Delta^2\mathbb{I}) \right)
\end{equation}
Fig.~\ref{fig:Efficient unlearning} presents the flowchart of the perturbed aggregation in each communication round (our contribution highlighted in orange). In storage critical applications where storing client weights at the server side is expensive, the server can only store the index of clients participating in each round of aggregation.
As soon as the server receives an unlearning request, the server removes the client based on its index from all the historical updates. In order for the server to plausibly unlearn the targeted client, each cluster must have at least $x (\leq k)$ number of model weights. Historically, if any cluster has fewer than $x$ model weights, the server rollbacks to the previous state, employs the unlearning mechanism, and recomputes the clusters. In case the server does not store the client gradients, the unlearning mechanism has to be exact unlearning i.e., retrain the models from that state onwards. 


\textbf{PoD for $\mathcal{SG}_{FL}$:} Consider a scenario where a client $c^*$ requests central server to unlearn. The central server removes the index of $c^*$ from all the stored clusters and if they have more than $x$ number of client weights, then under plausible deniability the server does not need to employ an unlearning mechanism. Now, we know that the model updates in each cluster is at most $\Delta$ distant apart. We know from the convergence analysis in~\cite{varshneyk} that $\Delta$ is a small value and under Conjecture 1 we can say that the models in each cluster are functionally equivalent with respect to MIA. In this case, even if the adversary ($\mathcal{A}$) in the security game $\mathcal{SG}_{FL}()$ predicts 1, i.e., $c^*$ was part of the training, then the server can offer a \textit{Proof of Deniability}, which comprises logs of model updates or indices from \(\mathbb{C} \setminus c^*\) that lead to a similar global model. Hence, the server can ($x, \Delta$) plausibly deny the participation of the targeted client $c^*$ in training.

\subsection{Client-level Privacy Analysis}

In our methodology, we randomly chose a cluster representative in each round and perturb it in order to avoid any inference by the central server, i.e., even the central server can not know which model weights has been used for aggregation. Now, we provide the privacy analysis of our approach. Let us consider $\mathcal{W}$ and $\mathcal{W}'$ be the set of weights in a cluster (say $C_c$) such that $\mathcal{W} = \mathcal{W}' \cup x^*$, and $x^*$ be a client weight. Then from the definition of $l_2-sensitivity$ (ref.~\ref{sec: diff priv}), $l_2-sensitivity$ of the model updates in a cluster is given by:
\begin{align}
    \psi(C_c) = \max_{w_i \in \mathcal{W}, w_j \in \mathcal{W}'} ||w_i \ominus w_j ||
\end{align}
We know that the model updates in a cluster are in the radius of $\Delta$ and two models can not differ by more than $2\Delta$ and therefore, $\psi(C_c) \leq 2\Delta$. Let us now consider the Gaussian perturbation defined by:
\begin{eqnarray}
GM = w_{r_c} + N (0, \sigma^2\psi(C_c)^2\mathbb{I})
   &=& w_{r_c} + N (0, 4\sigma^2\Delta^2\mathbb{I}) \nonumber \\
   &=& w_{r_c} + N (0, (2\sigma)^2\Delta^2\mathbb{I})
\end{eqnarray}
where $r_c$ is the index of randomly selected weight in cluster $C_c$. Now, we know from Lemma 2, $GM$ satisfies $\left( \alpha, \frac{\alpha}{8\sigma^2} \right)$-RDP i.e., in a given communication round, each cluster in our methodology satisfies $\left( \alpha, \frac{\alpha}{8\sigma^2} \right)$-RDP. 

Since the GM satisfies $\left( \alpha, \frac{\alpha}{8\sigma^2} \right)$-RDP independently in all the clusters, the aggregated global model $(w_{t}^g)$ at communication round $t$ is also $(\alpha, \frac{\alpha}{8\sigma^2} )$-RDP protected. Using Lemma 3, after $T$ iterations, the global model $w^g$ is $( \alpha, \frac{T\alpha}{8\sigma^2} )$-RDP protected. Now, in order to guarantee $(\epsilon, \delta)$-DP we use Lemma 1 to get the inequality,
\begin{align}
    \frac{T\alpha}{8\sigma^2} + \frac{\log(1/\delta)}{\alpha -1} \leq \epsilon
\end{align}
Suppose we choose $\alpha = 1 + 8\log(1/\delta)/\epsilon$, then we have
\begin{align}
    \sigma^2 \geq \frac{T(1+8\log(1/\delta))}{7\epsilon^2}
\end{align}
Based on this result, we establish the following theorem.

\begin{theorem}
    Given $0< \epsilon < 8\log(1/\delta)$ and $\delta>0$, the global model $w^g$ satisfies $(\epsilon, \delta)$-differential privacy after $T$ communication rounds iff
    \begin{align}
        \sigma^2 \geq \frac{T(1+8\log(1/\delta))}{7\epsilon^2} 
    \end{align}
\end{theorem}

\section{Experimental Analysis}

\begin{table}[htbp] \label{Experimental setup}
    \floatconts 
    {tab:experimental setup}
    {\caption{Details of the experimental setup.}}
    {%
        \begin{tabular}{|c|c|c|}
        \hline
        Parameters & Values & Description  \\ \hline
        Clients & 50 & \makecell{Number of clients in each \\ round of communication} \\ \hline
        Global Server & 1 & Server aggregate the local models \\ \hline
        Algorithms compared & 3 & fedAvg\_retrain, $k$-IPfedAvg, fedEraser \\ \hline
        $k$ in $k$-IPfedAvg & 4,6,8,10 & \makecell{Determines the number of \\ clients in each cluster} \\ \hline
        x in plausible deniability & 2,3,4 & \makecell{Determines the amount of noise \\ needed while training}\\ \hline
        Datasets & \makecell{MNIST, \\ CIFAR10, CelebA, \\FashionMNIST} & \makecell{iid and non-iid distribution\\ of these datasets} \\ \hline
        Local Epochs & 3 & \makecell{Number of local training \\ iterations in each round} \\ \hline
        Global rounds & 50 & \makecell{Number of communications \\ between server and uses.} \\ \hline
        unlearning probability & 0.2 & \makecell{In each communication, a client reque\\-sts for unlearning with probability 0.2.} \\ \hline
    \end{tabular}
    }
\end{table}

In this section, we present the experimental setup and analysis of our proposed methodology. In this work, we have simulated the FL environment on a local machine. We have created 50 clients and they train the global model for 50 communication rounds. In a given round of communication, each user trains the global model for 3 epochs on their local data and then communicates its model updates back to the server. Table~\ref{tab:experimental setup} provides the details of the experimental setup. In our work, we consider unlearning can be requested throughout the training rounds while most of the work in the literature considers 1 unlearning request in their experimental setup. We have considered three different network architectures to show that our methodology has good performance on a variety of CNNs. We have experimented with a custom CNN (ConvNet from now onwards) which consists of two convolution layers (first layer with 20 filters, second layer with 10 filters with $(3,3)$ as kernel size) and a dense layer (32 neurons) as hidden layers, LeNet5 (\cite{lecun1998gradient}), and a custom residual network (ResNet-mini from now onwards) with two residual blocks connected to a fully connected layer with total 7 layers of learnable parameters ($\approx$ ResNet-7) . We compare our methodology with fedAvg (\cite{mcmahan2017communication}) and fedEraser (\cite{liu2021federaser}) to evaluate the performance of our method. 

We have validated our approach using four datasets: MNIST (\cite{deng2012mnist}), which consists of 60,000 images for training and validation and 10,000 for testing; FashionMNIST (\cite{xiao2017fashion}), with the same image distribution as MNIST; CIFAR10 (\cite{krizhevsky2012imagenet}), which comprises 50,000 training and validation images and 10,000 testing images; and CelebA (\cite{liu2015faceattributes}) consists of more than 200K celebrity images with 40 attribute labels. For training, 50K samples were selected from the original 163K training images, and 20K samples were used as the test set. The MNIST, FashionMNIST, and CIFAR10 datasets each have ten output classes, while the CelebA dataset is a multi-label classification dataset with 40 binary attribute labels. They have been analyzed in the identically distributed (iid) and non-independent and identically distributed (non-iid) manner to validate the performance in heterogeneous FL setting.

\begin{figure}[ht]
    \floatconts
        {Test_acc scores} 
        {\caption{The test accuracy (y-axis) of the ConvNet model for $k$-IPfedAvg several values of $k$ (4,6,8,10) and several degree of deniability ($x$: {2,3,4}), along with fedAvg, fedEraser during the unlearning process for: (a) MNIST-iid (b) FashionMNIST-iid (c) CIFAR10-iid (d) CelebA-iid (e) MNIST-noniid (f) FashionMNIST-noniid (g) CIFAR10-noniid (h) CelebA-noniid}}
        {
            \subfigure{
                \includegraphics[width=0.22\textwidth]{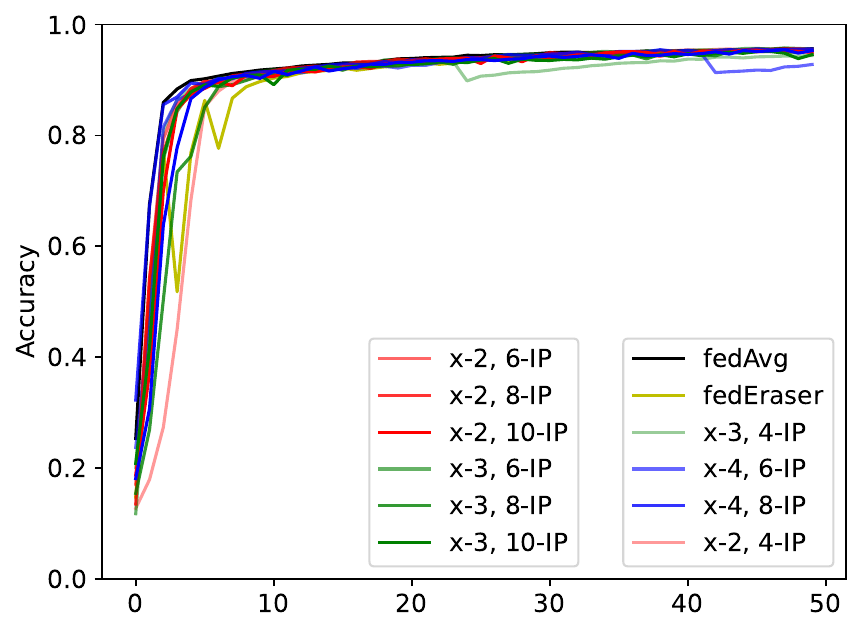}
                \label{MNISTiidtest_acc}
            }
            \subfigure{
                \includegraphics[width=0.22\textwidth]{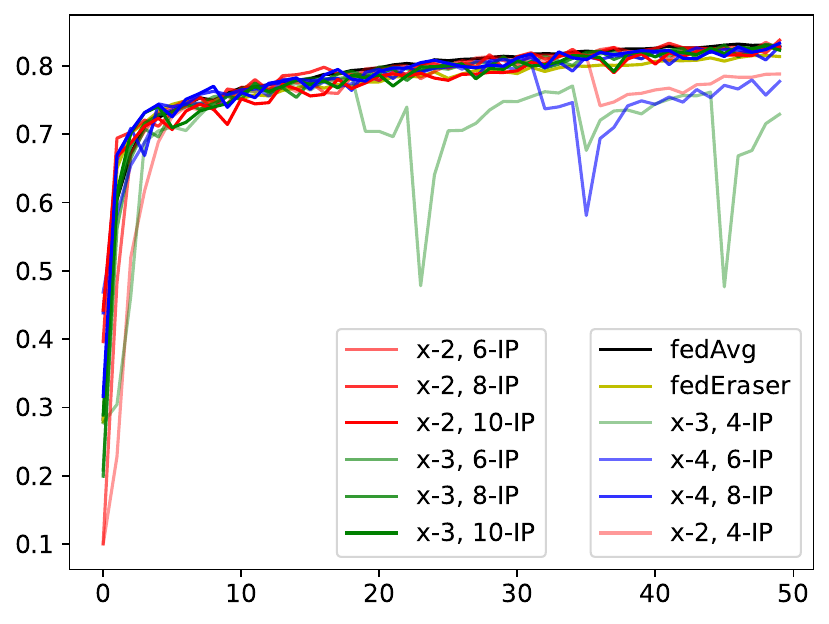}
                \label{FMNISTiidtest_acc}
            }
            \subfigure{
                \includegraphics[width=0.22\textwidth]{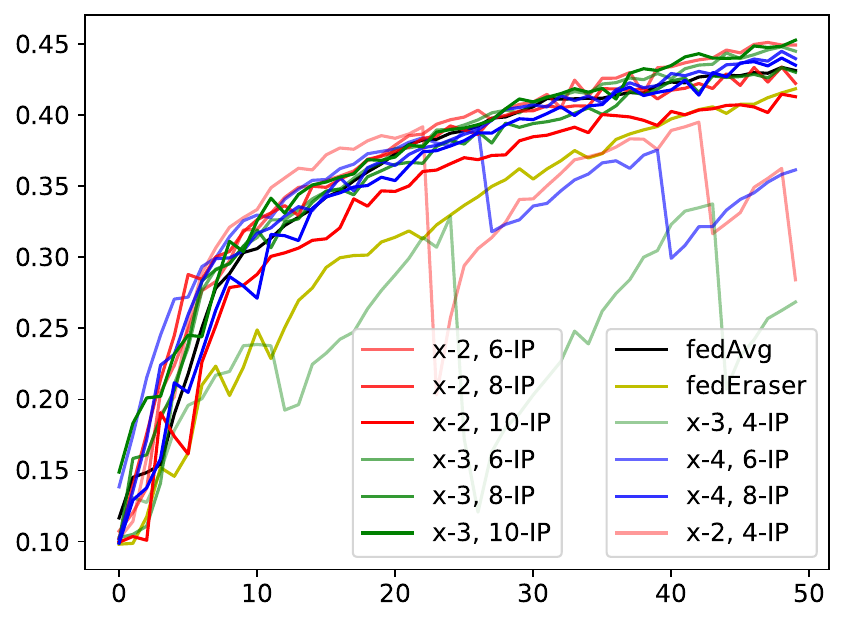}
                \label{CIFAR10iidtest_acc}
            }
            \subfigure{
                \includegraphics[width=0.22\textwidth]{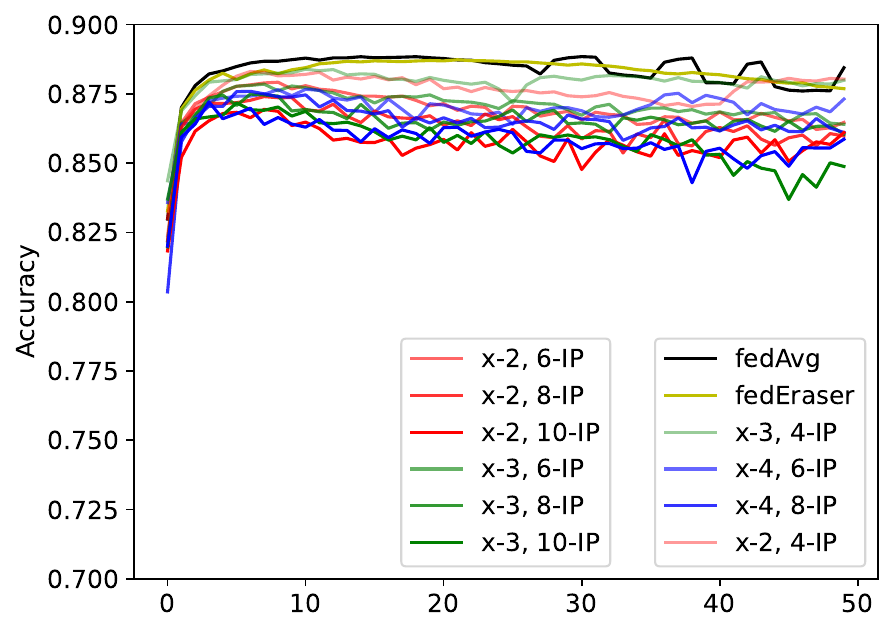}
                \label{celebAiidtest_acc}
            }
            
            \subfigure{
                \includegraphics[width=0.22\textwidth]{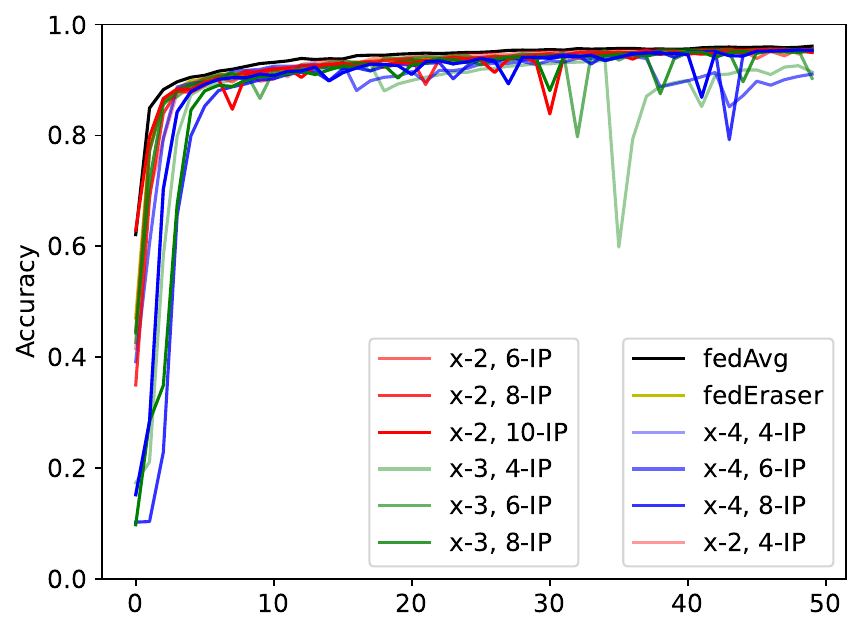}
                \label{MNISTnoniidtest_acc}
            }
            \subfigure{
                \includegraphics[width=0.22\textwidth]{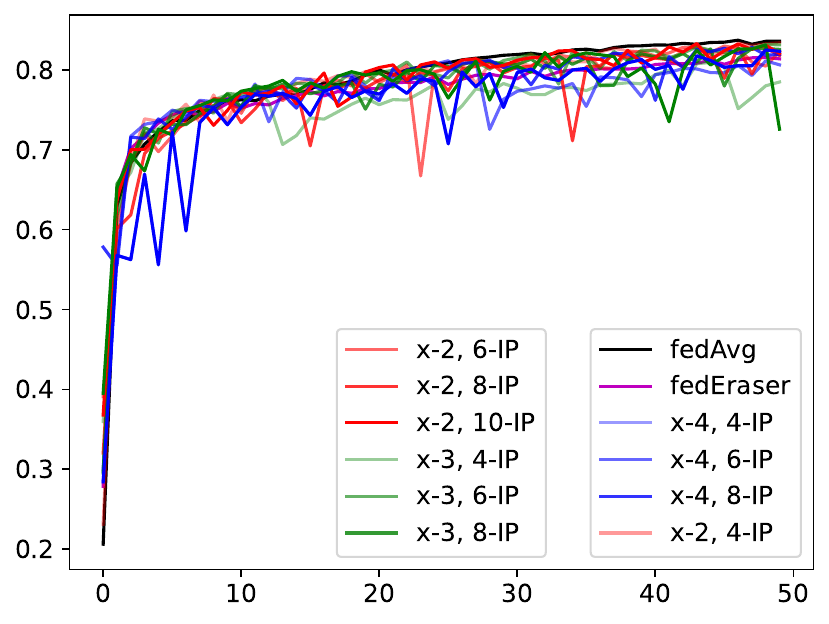}
                \label{FMNISTnoniidtest_acc}
            }
            \subfigure{
                \includegraphics[width=0.22\textwidth]{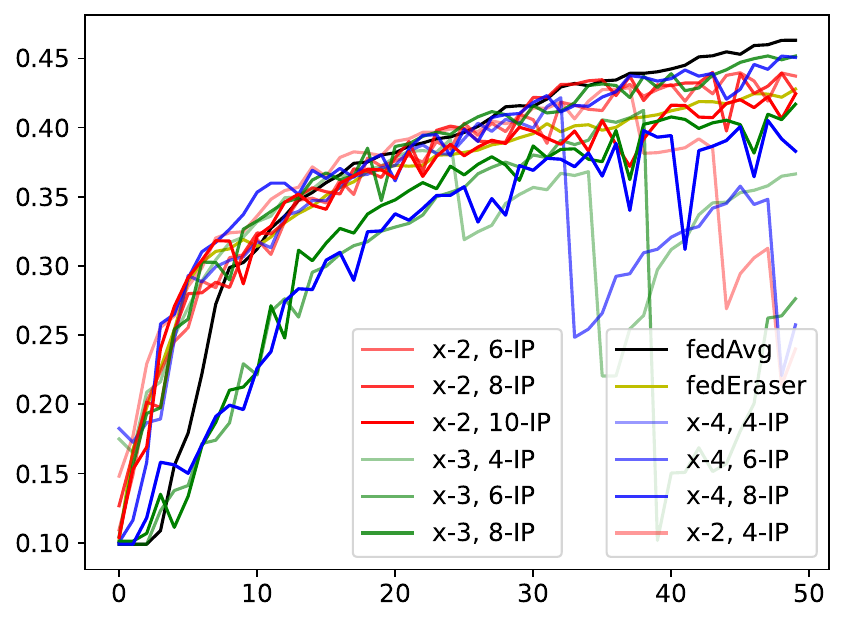}
                \label{CIFAR10noniidtest_acc}
            }
            \subfigure{
                \includegraphics[width=0.22\textwidth]{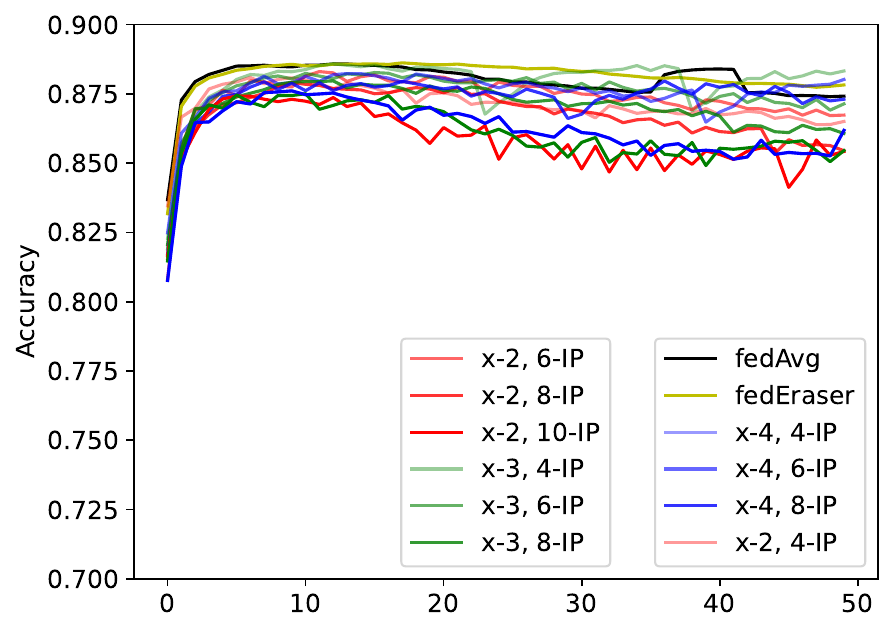}
                \label{celebAnoniidtest_acc}
            }

        }
\end{figure}

\begin{table}[htbp]
    \centering 
    \caption{Comparison of test accuracy between fedAvg, fedEraser and perturbed $k$-IPfedAvg ($k=4,6,8,10$) for LeNet5. The results for perturbed $k$-IPfedAvg shows the mean along with its standard deviation for $x=2,3,4$.}
    \scalebox{0.9}{
    \begin{tabular}{|l|l|l|l|l|l|l|}
        \hline
        \multirow{2}{*}{Dataset} & \multirow{2}{*}{fedAvg} & \multirow{2}{*}{fedEraser} & \multicolumn{4}{c|}{Perturbed $k$-IPfedAvg} \\ \cmidrule{4-7}
        & & & k=4 & k=6 & k=8 & k=10 \\
        \hline
        MNIST-iid & $0.964$ & $0.947$ &  $0.927 \pm 0.02$ & $0.922 \pm 0.03$ & $0.961 \pm 0.01$ & $0.964 \pm 0.0$ \\ \hline
        MNIST-noniid & $0.968$ & $0.958$ & $0.943 \pm 0.02$ & $0.925 \pm 0.05$ & $0.961 \pm 0.0$ & $0.963 \pm 0.0$ \\ \hline
        FMNIST-iid & $0.794$ & $0.756$ & $0.635 \pm 0.02$ & $0.777 \pm 0.01$ & $0.784 \pm 0.0$ & $0.777 \pm 0.01$ \\ \hline
        FMNIST-noniid & $0.801$ & $0.757$ & $0.75 \pm 0.01$ & $0.763 \pm 0.04$ & $0.785 \pm 0.01$ & $0.778 \pm 0.01$ \\ \hline
        CIFAR10-iid & $0.411$ & $0.301$ & $0.223 \pm 0.04$ & $0.355 \pm 0.03$ & $0.389 \pm 0.02$ & $0.319 \pm 0.01$ \\ \hline
        CIFAR10-noniid & $0.438$ & $0.319$ & $0.321 \pm 0.05$ & $0.276 \pm 0.06$ & $0.39 \pm 0.01$ & $0.396 \pm 0.01$ \\ \hline
        CelebA-iid & $0.862$ & $0.839$ & $0.808 \pm 0.02$ & $0.823 \pm 0.01$ & $0.832 \pm 0.01$ & $0.817 \pm 0.01$ \\ \hline 
        CelebA-noniid & $0.859$ & $0.831$ & $0.804 \pm 0.0$ & $0.802 \pm 0.02$ & $0.827 \pm 0.03$ & $0.816 \pm 0.02$ \\ \hline

    \end{tabular}}
    \label{tab:Lenet_acc}
\end{table}

\begin{table}[h]
    \centering 
    \caption{Comparison of test accuracy between fedAvg, fedEraser and perturbed $k$-IPfedAvg ($k=4,6,8,10$) for ResNet-mini. The results for perturbed $k$-IPfedAvg shows the mean along with its standard deviation for $x=2,3,4$.}
    \scalebox{0.9}{
    \begin{tabular}{|l|l|l|l|l|l|l|}
        \hline
        \multirow{2}{*}{Dataset} & \multirow{2}{*}{fedAvg} & \multirow{2}{*}{fedEraser} & \multicolumn{4}{c|}{Perturbed $k$-IPfedAvg} \\ \cline{4-7}
        & & & k=4 & k=6 & k=8 & k=10 \\
        \hline
        MNIST-iid & $0.977$ & $0.963$ &  $0.963 \pm 0.01$ & $0.972 \pm 0.0$ & $0.974 \pm 0.0$ & $0.972 \pm 0.0$ \\ \hline
        MNIST-noniid & $0.975$ & $0.97$ & $0.945 \pm 0.01$ & $0.942 \pm 0.03$ & $0.973 \pm 0.0$ & $0.971 \pm 0.0$ \\ \hline
        FMNIST-iid & $0.871$ & $0.857$ & $0.831 \pm 0.04$ & $0.869 \pm 0.0$ & $0.868 \pm 0.0$ & $0.868 \pm 0.0$ \\ \hline
        FMNIST-noniid & $0.879$ & $0.861$ & $0.865 \pm 0.01$ & $0.85 \pm 0.01$ & $0.854 \pm 0.02$ & $0.863 \pm 0.01$ \\ \hline
        CIFAR10-iid & $0.543$ & $0.526$ & $0.478 \pm 0.02$ & $0.516 \pm 0.02$ & $0.529 \pm 0.01$ & $0.517 \pm 0.01$ \\ \hline
        CIFAR10-noniid & $0.549$ & $0.536$ & $0.459 \pm 0.0$ & $0.507 \pm 0.03$ & $0.51 \pm 0.01$ & $0.511 \pm 0.02$ \\ \hline
        CelebA-iid & $0.877$ & $0.869$ & $0.865 \pm 0.0$ & $0.866 \pm 0.01$ & $0.868 \pm 0.0$ & $0.867 \pm 0.0$ \\ \hline 
        CelebA-noniid & $0.878$ & $0.869$ & $0.865 \pm 0.0$ & $0.863 \pm 0.01$ & $0.867 \pm 0.0$ & $0.862 \pm 0.01$ \\ \hline
    \end{tabular}}
    \label{tab:resnet_acc}
\end{table}


\begin{table}[!h]
    \centering 
    \caption{Comparison of the average wall-clock running time (in seconds) between fedAvg, and perturbed $k$-IPfedAvg ($k=4,6,8,10$) for LeNet5. The results for perturbed $k$-IPfedAvg shows the mean along with its standard deviation for $x=2,3,4$.}
    \begin{tabular}{|l|l|l|l|l|l|}
        \hline
        \multirow{2}{*}{Dataset} & \multirow{2}{*}{fedAvg} & \multicolumn{4}{c|}{Perturbed $k$-IPfedAvg} \\ \cline{3-6}
        & & k=4 & k=6 & k=8 & k=10 \\
        \hline
        MNIST-iid & $225.61$ &  $101.5 \pm 101$ & $12.71 \pm 17.9$ & $0.021 \pm 0.01$ & $0.024 \pm 0.0$ \\ \hline
        MNIST-noniid & $260.14$ & $40.5 \pm 27.6$ & $20.89 \pm 19.1$ & $0.021 \pm 0.0$ & $0.02 \pm 0.0$ \\ \hline
        FMNIST-iid & $217.02$ & $114.9 \pm 54.4$ & $7.14 \pm 10.1$ & $0.022 \pm 0.0$ & $0.022 \pm 0.0$ \\ \hline
        FMNIST-noniid & $223.77$ & $56.6 \pm 26.2$ & $12.78 \pm 18$ & $0.023 \pm 0.0$ & $0.025 \pm 0.0$ \\ \hline
        CIFAR10-iid & $294.78$ & $41.78 \pm 9.19$ & $7.91 \pm 11$ & $0.018 \pm 0.0$ & $0.021 \pm 0.0$ \\ \hline
        CIFAR10-noniid & $295.53$ & $15.46 \pm 15$ & $0.024 \pm 0.0$ & $0.026 \pm 0.0$ & $0.022 \pm 0.0$ \\ \hline 
        CelebA-iid & $1362.2$ & $225.5 \pm 319$ & $105.3 \pm 182.4$ & $0.004 \pm 0.0$ & $0.006 \pm 0.0$ \\ \hline 
        CelebA-noniid & $1270.3$ & $382.95 \pm 72.7$ & $150.6 \pm 134.8$ & $52.6 \pm 91.2$ & $0.005 \pm 0.0$ \\ \hline

    \end{tabular}
    \label{tab:Lenet_running time.}
\end{table}

\begin{table}[!h]
    \centering 
    \caption{Comparison of the average wall-clock running time (in seconds) between fedAvg, and perturbed $k$-IPfedAvg ($k=4,6,8,10$) for ResNet-mini. The results for perturbed $k$-IPfedAvg shows the mean along with its standard deviation for $x=2,3,4$.}
    \begin{tabular}{|l|l|l|l|l|l|}
        \hline
        \multirow{2}{*}{Dataset} & \multirow{2}{*}{fedAvg} & \multicolumn{4}{c|}{Perturbed $k$-IPfedAvg} \\ \cline{3-6}
        & & k=4 & k=6 & k=8 & k=10 \\
        \hline
        MNIST-iid & $354.37$ &  $94.34 \pm 59.83$ & $104.33 \pm 147.5$ & $0.005 \pm 0.0$ & $0.017 \pm 0.0$ \\ \hline
        MNIST-noniid & $302.34$ & $235.44 \pm 95.48$ & $207.1 \pm 155$ & $0.016 \pm 0.01$ & $0.012 \pm 0.0$ \\ \hline
        FMNIST-iid & $253.09$ & $206.43 \pm 206$ & $0.002 \pm 0.0$ & $0.007 \pm 0.0$ & $0.007 \pm 0.0$ \\ \hline
        FMNIST-noniid & $331.2$ & $141.4 \pm 141$ & $124.69 \pm 90.9$ & $14.56 \pm 20.6$ & $0.005 \pm 0.0$ \\ \hline
        CIFAR10-iid & $286.37$ & $166.3 \pm 75.64$ & $44.8 \pm 63.4$ & $0.007 \pm 0.0$ & $0.007 \pm 0.0$ \\ \hline
        CIFAR10-noniid & $311.28$ & $116.9 \pm 16$ & $46.37 \pm 65.57$ & $16.78 \pm 23.7$ & $0.004 \pm 0.0$ \\ \hline 
        CelebA-iid & $1084.9$ & $572 \pm 435$ & $425 \pm 409.4$ & $104.7 \pm 181.2$ & $0.033 \pm 0.0$ \\ \hline 
        CelebA-noniid & $577.2$ & $442.6 \pm 104.3$ & $100.2 \pm 134.8$ & $0.013 \pm 0.0$ & $0.01 \pm 0.0$ \\ \hline
    \end{tabular}
    \label{tab:resnet_running time.}
\end{table}

\begin{figure}[ht]
    \floatconts
    {Comparison of disk space} 
    {\caption{The comparison of Disk Space (y-axis) for $k$-IPfedAvg, and fedEraser for: (a) MNIST-ConvNet (b) FashionMNIST-ConvNet (c) CIFAR10-ConvNet (d) CelebA-ConvNet}} 
    {%
        \subfigure{%
            \label{MNISTiid_disk_space-ConvNet}
            \includegraphics[width=0.23\textwidth]{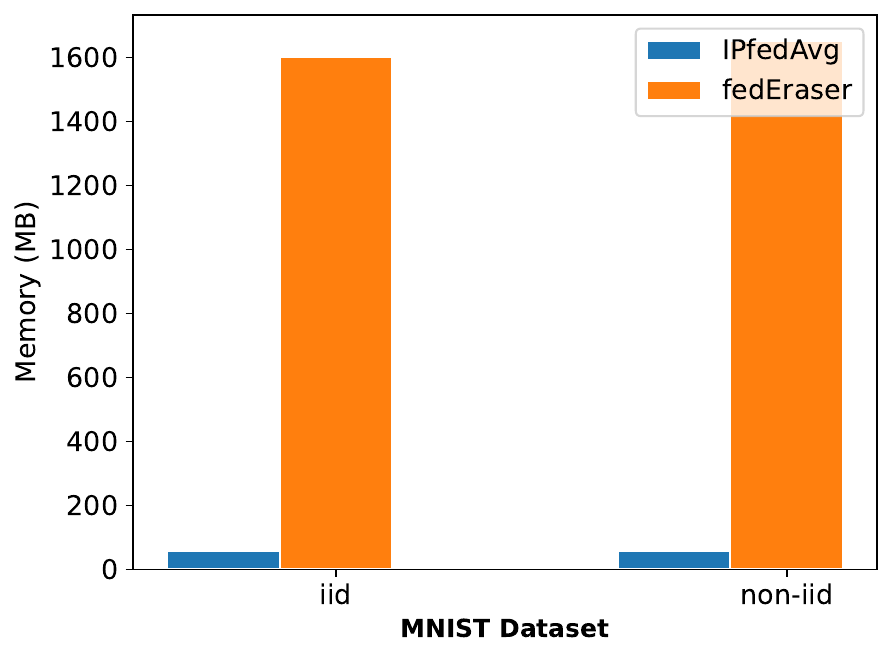}
        }
        \subfigure{%
            \label{FMNISTiid_disk_space-ConvNet}
            \includegraphics[width=0.23\textwidth]{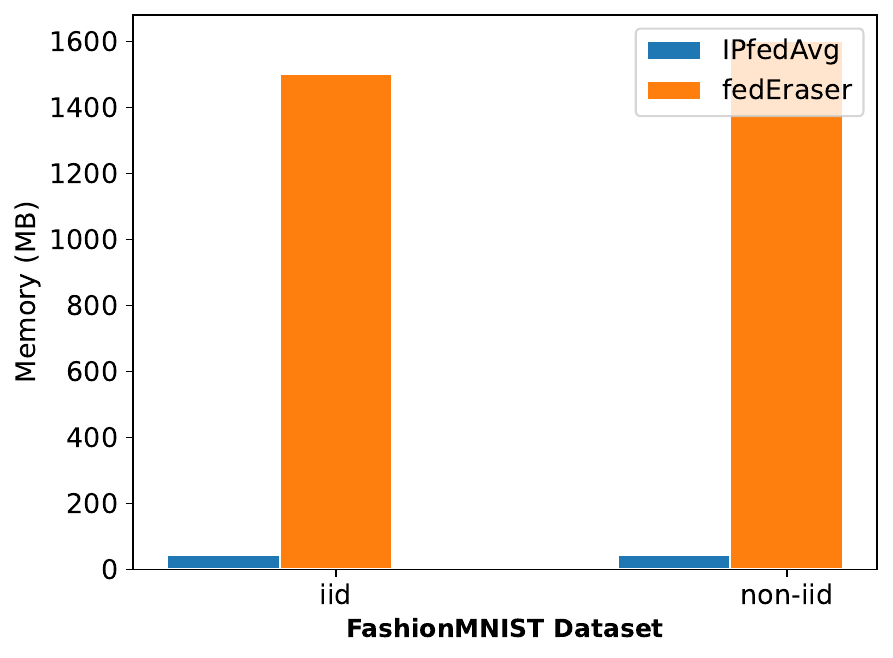}
        }
        \subfigure{%
            \label{CIFAR10iid_disk_space-ConvNet}
            \includegraphics[width=0.23\textwidth]{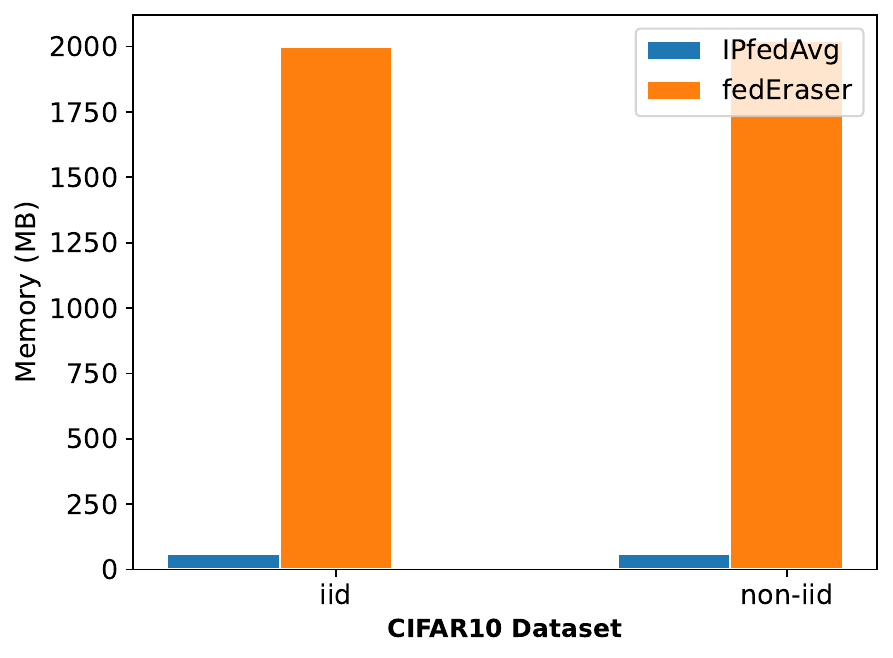}
        }
        \subfigure{%
            \label{celebA_diskSpace}
            \includegraphics[width=0.23\textwidth]{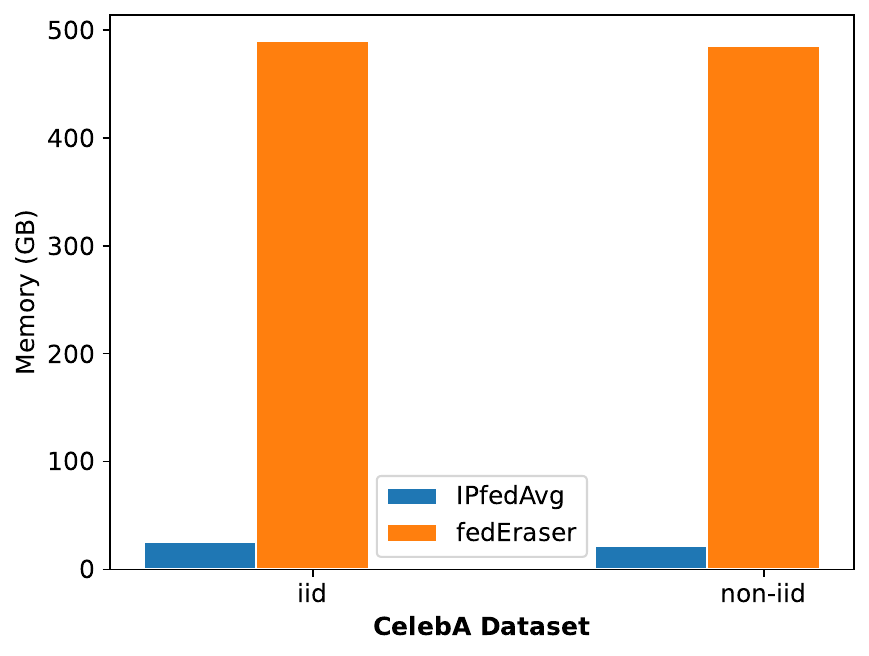}
}

    }
\end{figure}

In our experiments, for each communication round, a randomly selected client sends unlearning request to the server with a probability of $0.2$. Essentially, throughout the training, the expected number of unlearning requests to the server is 10. We consider retraining as the unlearning mechanism. The fedAvg with retraining is our benchmark; we also compare our methodology with fedEraser~\cite{liu2021federaser} which does not retrain the global model but uses historical model updates to compute the global model for each round. We use the $cosine$ distance measure to compute the distance between two models in Algorithm~\ref{Algo IPfedAvg}. 

Fig.~\ref{Test_acc scores} shows the comparison of test accuracy for ConvNet on the iid, and non-iid distributions of MNIST, FashionMNIST and CIFAR10 datasets. It shows that our proposed methodology has training accuracy comparable to federated learning, and fedEraser under retraining in most settings and improved in some. The results also show that our methodology offers various options for the privacy and deniability parameters where the test accuracy results are comparable. We also observe sudden drops in accuracy from Fig.~\ref{Test_acc scores} when the plausible deniability parameter is high (i.e., $x$ is high) and privacy parameter $k$ in $k$-IPfedAvg is low, making it a poor choice for selection. The reason for low test accuracy can be the higher number of retraining, and hence the sudden accuracy drop. It is very interesting to see here that with high $k$ values, i.e., by employing a better privacy mechanism during training, we still have benchmark comparable accuracy in all the cases. We also present the test accuracy results of LeNet5 in Table~\ref{tab:Lenet_acc}, and ResNet-mini in Table~\ref{tab:resnet_acc}. The findings indicate that perturbed $k$-IPfedAvg achieves benchmark-comparable results, particularly for higher values of $k$. Interestingly, as $k$ increases, the standard deviation for different $x$ decreases, suggesting no (or infrequent) retraining even for stronger plausible deniability parameter.


Table~\ref{tab:Lenet_running time.} and Table~\ref{tab:resnet_running time.} present the comparison of running wall-clock time between our methodology and fedAvg with retraining for LeNet5 and ResNet-mini respectively. The results clearly demonstrate that our methodology achieves at least $1.6\times$ improvement compared to fedAvg with retraining for small $k$ values. Interestingly, when the privacy parameter is set to higher values (such as $8$ or $10$),  no retraining is required (resulting in up to a staggering 500769 $\times$ improvement with $k=8$ for FMNIST in noniid setting). This suggests the negative correlation between the privacy parameter and the retraining time. Similar trend was observed for ConvNet model (see Appendix Fig.~\ref{unlearning time} ), where increasing the value of k has a clear impact on both the retraining time and the  deviation across different parameters.

Our methodology also saves on disk storage (for storage critical applications) at the server side in comparison with approximate unlearning for federated learning. Here, the influence of the targeted client is approximately computed and systematically removed (as in fedEraser~\cite{liu2021federaser}) in Fig.~\ref{Comparison of disk space} for ConvNet. The figure illustrates the significant storage improvement achieved with our framework ($\approx$ 30 times better in all cases). Specifically, we only need to store the client ID during clustering, whereas fedEraser stores the client updates in each communication round. Similar trend was observed for LeNet5 and ResNet-mini as well (see Fig.~\ref{Comparison of add disk space} in Appendix).



\section{Conclusion and Future Works}

In this paper, we have presented a novel plausible deniable framework for federated unlearning which reduces the need to employ unlearning mechanism by the server significantly. In our work, the server clusters the client's weight based on some distance measure and randomly picks a client from each cluster and perturbs it to avoid inference. The perturbation is necessary to avoid any inference by the honest but curious server. For every unlearning request, the server removes the client id from the cluster in all communication round. To avoid employing unlearning mechanism (retraining in our case), the server ensures it has at least $x$ number of clients in each cluster in all the historical updates, if not it rolls back to the previous round and employs an unlearning mechanism. The flexibility of plausible deniability allows the server to reduce the number of retraining requests. We also show that after $T$ number of communication rounds, the global model is $(\epsilon, \delta)$-differentially private for $0< \epsilon < 8\log(1/\delta)$ and $\delta>0$. Our approach reduces the number of retraining and disk storage for the server during federated unlearning. For future work, we plan to consider unlearning requests in large language models and generative models. Furthermore, determining the plausible deniability parameter in unlearning can be application dependent. A comprehensive examination of how plausible deniable unlearning aligns with AI regulations such as GDPR also presents an interesting direction.

\vspace{0.2cm}

\bibliography{acml24}

\begin{thebibliography}{23}
\providecommand{\natexlab}[1]{#1}
\providecommand{\url}[1]{\texttt{#1}}
\expandafter\ifx\csname urlstyle\endcsname\relax
  \providecommand{\doi}[1]{doi: #1}\else
  \providecommand{\doi}{doi: \begingroup \urlstyle{rm}\Url}\fi

\bibitem[Bindschaedler et~al.(2017)Bindschaedler, Shokri, and Gunter]{bindschaedler2017plausible}
Vincent Bindschaedler, Reza Shokri, and Carl~A Gunter.
\newblock Plausible deniability for privacy-preserving data synthesis.
\newblock \emph{arXiv preprint arXiv:1708.07975}, 2017.

\bibitem[Carlini et~al.(2022)Carlini, Chien, Nasr, Song, Terzis, and Tramer]{carlini2022membership}
Nicholas Carlini, Steve Chien, Milad Nasr, Shuang Song, Andreas Terzis, and Florian Tramer.
\newblock Membership inference attacks from first principles.
\newblock In \emph{2022 IEEE Symposium on Security and Privacy (SP)}, pages 1897--1914. IEEE, 2022.

\bibitem[Deng(2012)]{deng2012mnist}
Li~Deng.
\newblock The mnist database of handwritten digit images for machine learning research [best of the web].
\newblock \emph{IEEE signal processing magazine}, 29\penalty0 (6):\penalty0 141--142, 2012.

\bibitem[Geyer et~al.(2017)Geyer, Klein, and Nabi]{geyer2017differentially}
Robin~C Geyer, Tassilo Klein, and Moin Nabi.
\newblock Differentially private federated learning: A client level perspective.
\newblock \emph{arXiv preprint arXiv:1712.07557}, 2017.

\bibitem[Ginart et~al.(2019)Ginart, Guan, Valiant, and Zou]{ginart2019making}
Antonio Ginart, Melody Guan, Gregory Valiant, and James~Y Zou.
\newblock Making ai forget you: Data deletion in machine learning.
\newblock \emph{Advances in neural information processing systems}, 32, 2019.

\bibitem[Halimi et~al.(2022)Halimi, Kadhe, Rawat, and Baracaldo]{halimi2022federated}
Anisa Halimi, Swanand Kadhe, Ambrish Rawat, and Nathalie Baracaldo.
\newblock Federated unlearning: How to efficiently erase a client in fl?
\newblock \emph{arXiv preprint arXiv:2207.05521}, 2022.

\bibitem[Jayaraman et~al.(2020)Jayaraman, Wang, Knipmeyer, Gu, and Evans]{jayaraman2020revisiting}
Bargav Jayaraman, Lingxiao Wang, Katherine Knipmeyer, Quanquan Gu, and David Evans.
\newblock Revisiting membership inference under realistic assumptions.
\newblock \emph{arXiv preprint arXiv:2005.10881}, 2020.

\bibitem[Kong et~al.(2022)Kong, Roy~Chowdhury, and Chaudhuri]{kong2022forgeability}
Zhifeng Kong, Amrita Roy~Chowdhury, and Kamalika Chaudhuri.
\newblock Forgeability and membership inference attacks.
\newblock In \emph{Proceedings of the 15th ACM Workshop on Artificial Intelligence and Security}, pages 25--31, 2022.

\bibitem[Krizhevsky et~al.(2012)Krizhevsky, Sutskever, and Hinton]{krizhevsky2012imagenet}
Alex Krizhevsky, Ilya Sutskever, and Geoffrey~E Hinton.
\newblock Imagenet classification with deep convolutional neural networks.
\newblock \emph{Advances in neural information processing systems}, 25, 2012.

\bibitem[LeCun et~al.(1998)LeCun, Bottou, Bengio, and Haffner]{lecun1998gradient}
Yann LeCun, L{\'e}on Bottou, Yoshua Bengio, and Patrick Haffner.
\newblock Gradient-based learning applied to document recognition.
\newblock \emph{Proceedings of the IEEE}, 86\penalty0 (11):\penalty0 2278--2324, 1998.

\bibitem[Liu et~al.(2021)Liu, Ma, Yang, Wang, and Liu]{liu2021federaser}
Gaoyang Liu, Xiaoqiang Ma, Yang Yang, Chen Wang, and Jiangchuan Liu.
\newblock Federaser: Enabling efficient client-level data removal from federated learning models.
\newblock In \emph{2021 IEEE/ACM 29th International Symposium on Quality of Service (IWQOS)}, pages 1--10. IEEE, 2021.

\bibitem[Liu et~al.(2022)Liu, Xu, Yuan, Wang, and Li]{liu2022right}
Yi~Liu, Lei Xu, Xingliang Yuan, Cong Wang, and Bo~Li.
\newblock The right to be forgotten in federated learning: An efficient realization with rapid retraining.
\newblock In \emph{IEEE INFOCOM 2022-IEEE Conference on Computer Communications}, pages 1749--1758. IEEE, 2022.

\bibitem[Liu et~al.(2015)Liu, Luo, Wang, and Tang]{liu2015faceattributes}
Ziwei Liu, Ping Luo, Xiaogang Wang, and Xiaoou Tang.
\newblock Deep learning face attributes in the wild.
\newblock In \emph{Proceedings of International Conference on Computer Vision (ICCV)}, December 2015.

\bibitem[McMahan et~al.(2017)McMahan, Moore, Ramage, Hampson, and y~Arcas]{mcmahan2017communication}
Brendan McMahan, Eider Moore, Daniel Ramage, Seth Hampson, and Blaise~Aguera y~Arcas.
\newblock Communication-efficient learning of deep networks from decentralized data.
\newblock In \emph{Artificial intelligence and statistics}, pages 1273--1282. PMLR, 2017.

\bibitem[Mironov(2017)]{mironov2017renyi}
Ilya Mironov.
\newblock R{\'e}nyi differential privacy.
\newblock In \emph{2017 IEEE 30th computer security foundations symposium (CSF)}, pages 263--275. IEEE, 2017.

\bibitem[Suri et~al.(2022)Suri, Kanani, Marathe, and Peterson]{suri2022subject}
Anshuman Suri, Pallika Kanani, Virendra~J Marathe, and Daniel~W Peterson.
\newblock Subject membership inference attacks in federated learning.
\newblock \emph{arXiv preprint arXiv:2206.03317}, 2022.

\bibitem[Thudi et~al.(2022)Thudi, Jia, Shumailov, and Papernot]{thudi2022necessity}
Anvith Thudi, Hengrui Jia, Ilia Shumailov, and Nicolas Papernot.
\newblock On the necessity of auditable algorithmic definitions for machine unlearning.
\newblock In \emph{31st USENIX Security Symposium (USENIX Security 22)}, pages 4007--4022, 2022.

\bibitem[Torra et~al.(2020)Torra, Navarro-Arribas, and Galv{\'a}n]{torra2020explaining}
Vicen{\c{c}} Torra, Guillermo Navarro-Arribas, and Edgar Galv{\'a}n.
\newblock Explaining recurrent machine learning models: integral privacy revisited.
\newblock In \emph{International Conference on Privacy in Statistical Databases}, pages 62--73. Springer, 2020.

\bibitem[Varshney and Torra(2023{\natexlab{a}})]{varshney2023integrally}
A.K. Varshney and V.~Torra.
\newblock Integrally private model selection for deep neural networks.
\newblock \emph{Database and Expert Systems Applications. DEXA 2023}, 14147, 2023{\natexlab{a}}.

\bibitem[Varshney and Torra(2023{\natexlab{b}})]{varshneyk}
Ayush~K Varshney and Vicenc Torra.
\newblock k-ipfedavg: k-anonymous integrally private federated averaging with convergence guarantee.
\newblock \emph{techrxiv preprint 10.36227/techrxiv.170327604.45388443/v1}, 2023{\natexlab{b}}.

\bibitem[Wang et~al.(2023)Wang, Li, and Li]{wang2023federated}
Fei Wang, Baochun Li, and Bo~Li.
\newblock Federated unlearning and its privacy threats.
\newblock \emph{IEEE Network}, 2023.

\bibitem[Wu et~al.(2022)Wu, Zhu, and Mitra]{wu2022federated}
Chen Wu, Sencun Zhu, and Prasenjit Mitra.
\newblock Federated unlearning with knowledge distillation.
\newblock \emph{arXiv preprint arXiv:2201.09441}, 2022.

\bibitem[Xiao et~al.(2017)Xiao, Rasul, and Vollgraf]{xiao2017fashion}
Han Xiao, Kashif Rasul, and Roland Vollgraf.
\newblock Fashion-mnist: a novel image dataset for benchmarking machine learning algorithms.
\newblock \emph{arXiv preprint arXiv:1708.07747}, 2017.

\end{thebibliography}

\appendix

\section{Further Experiments}

\begin{figure}[ht]
    \floatconts
        {avg_dist scores} 
        {\caption{The plot for average distance between model generated by fedAvg and perturbed $k$-IPfedAvg for several ks ($4,6,8,10$): (a) MNIST-iid (b) FashionMNIST-iid (c) CIFAR10-iid (d) CelebA-iid (e) MNIST-noniid (f) FashionMNIST-noniid (g) CIFAR10-noniid (h) CelebA-noniid.}}
        {
            \subfigure{
                \includegraphics[width=0.22\textwidth]{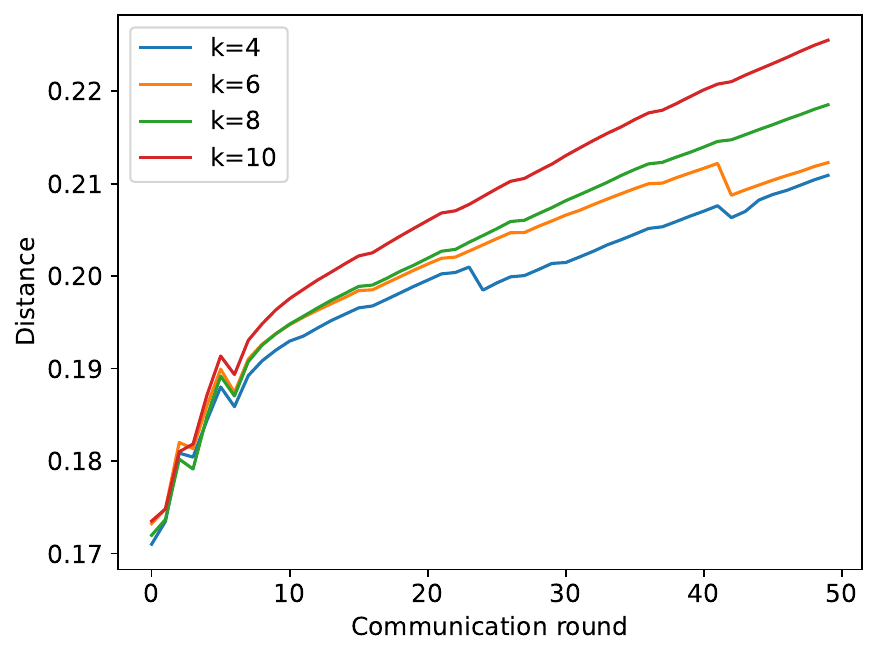}
                \label{MNISTiid_avg_dist}
            }
            \subfigure{
                \includegraphics[width=0.22\textwidth]{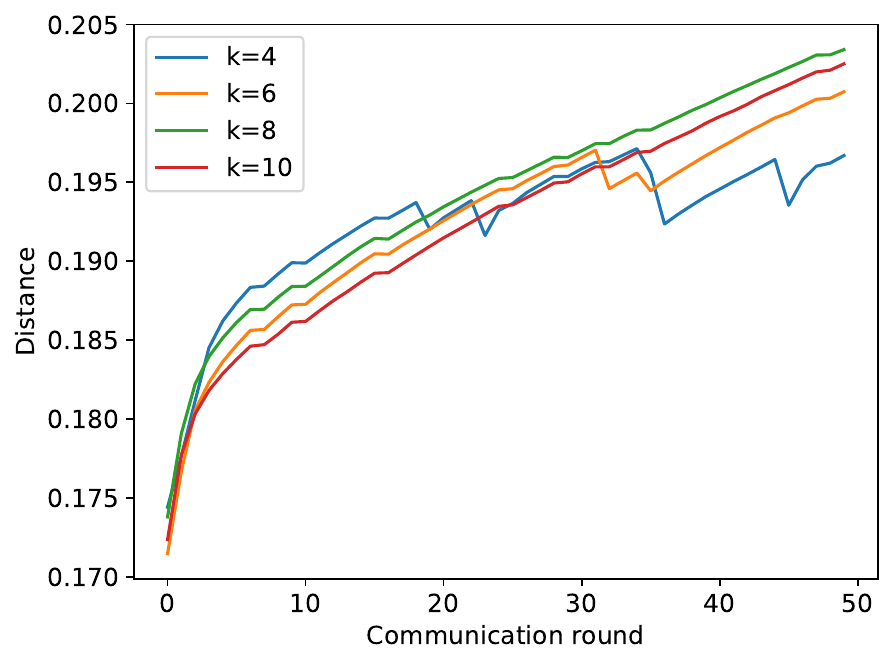}
                \label{FMNISTiid_avg_dist}
            }
            \subfigure{
                \includegraphics[width=0.22\textwidth]{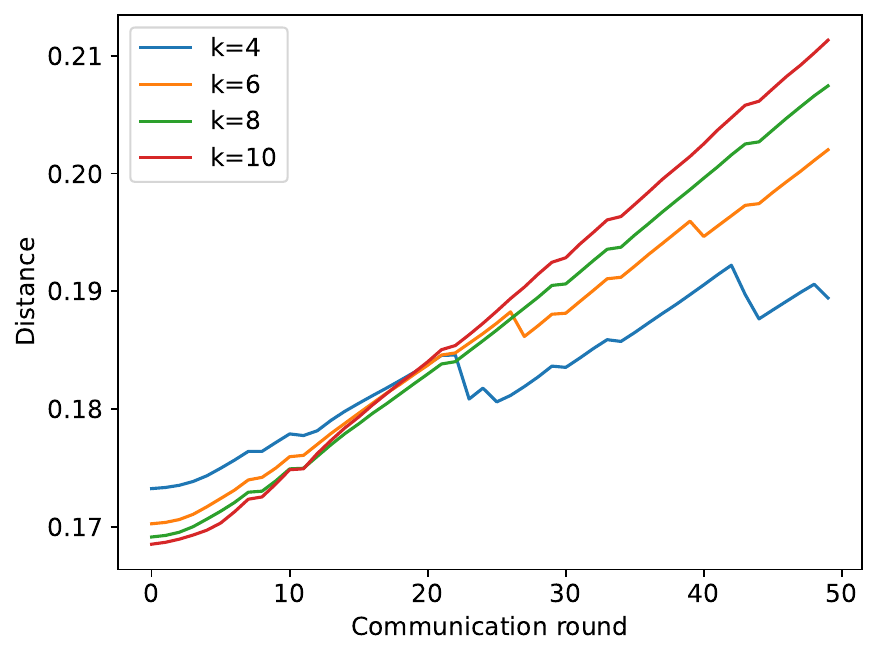}
                \label{CIFAR10iid_avg_dist}
            }
            \subfigure{
                \includegraphics[width=0.22\textwidth]{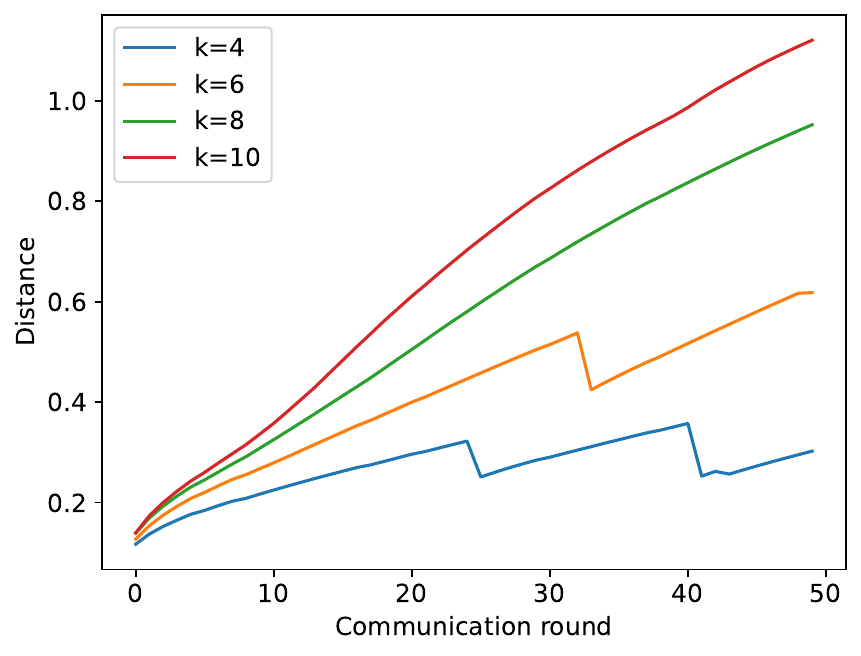}
                \label{celebAiid_avg_dist}
            }
            
            \subfigure{
                \includegraphics[width=0.22\textwidth]{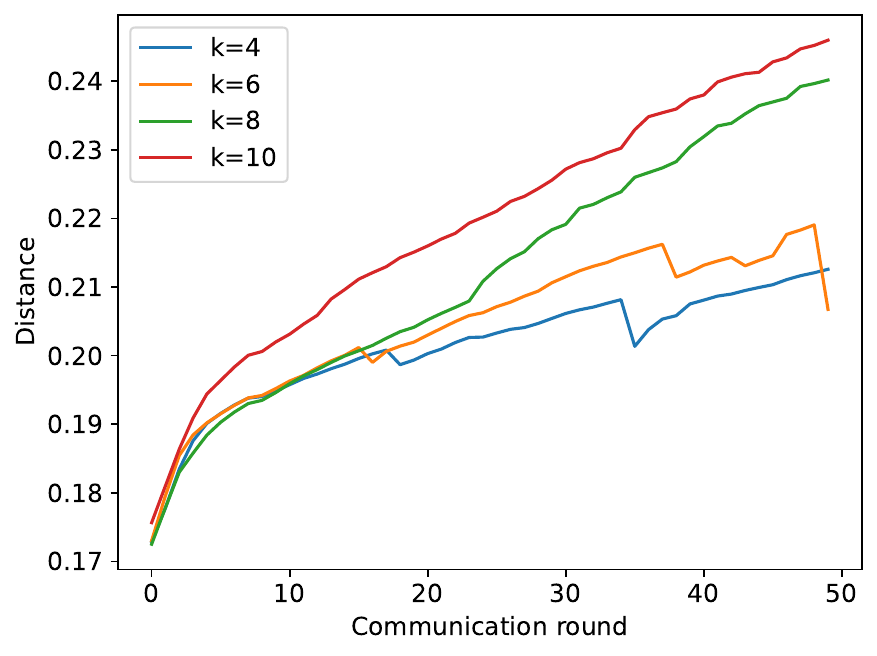}
                \label{MNISTnoniid_avg_dist}
            }
            \subfigure{
                \includegraphics[width=0.22\textwidth]{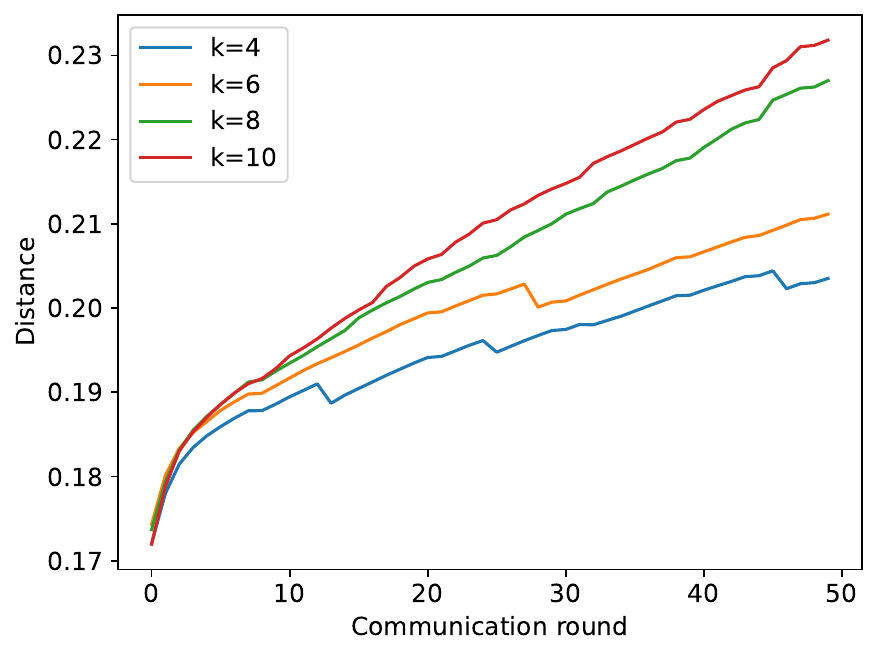}
                \label{FMNISTnoniid_avg_dist}
            }
            \subfigure{
                \includegraphics[width=0.22\textwidth]{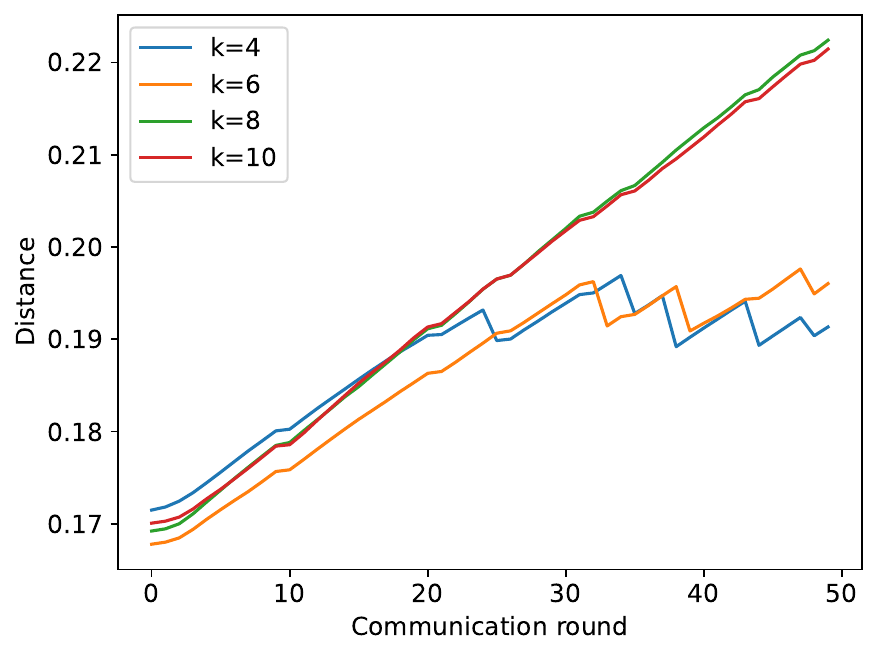}
                \label{CIFARnoniid_avg_dist}
            }
            \subfigure{
                \includegraphics[width=0.22\textwidth]{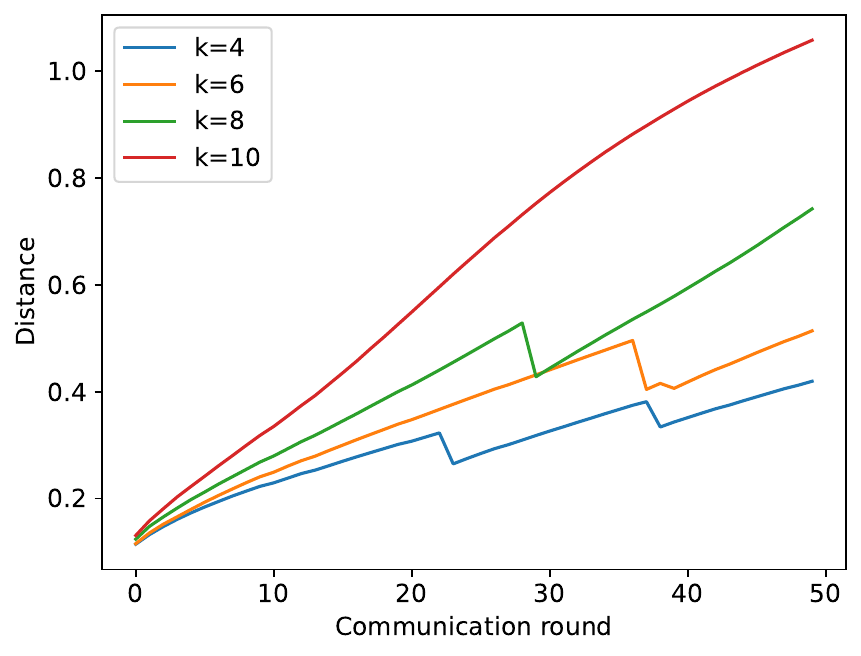}
                \label{celebAnoniid_avg_dist}
            }
        }
\end{figure}

Fig. \ref{avg_dist scores} highlights the average distance between the global model from perturbed $k$-IPfedAvg and fedAvg. As expected the distance gradually increases with communication round as the clients train on perturbed model in each communication round. However, for $k=4,6$ in Fig. \ref{avg_dist scores} the distance reduces due to higher retraining for lower $k$ values. 

\begin{figure}[ht]
    \floatconts
        {unlearning time} 
        {\caption{The comparison of unlearning time (y-axis) with ConvNet for $k$-IPfedAvg, and fedAvg for: (a) MNIST-iid (b) FashionMNIST-iid (c) CIFAR10-iid (d) CelebA-iid (e) MNIST-noniid (f) FashionMNIST-noniid (g) CIFAR10-noniid (h) CelebA-noniid.}}
        {
            \subfigure{
                \includegraphics[width=0.22\textwidth]{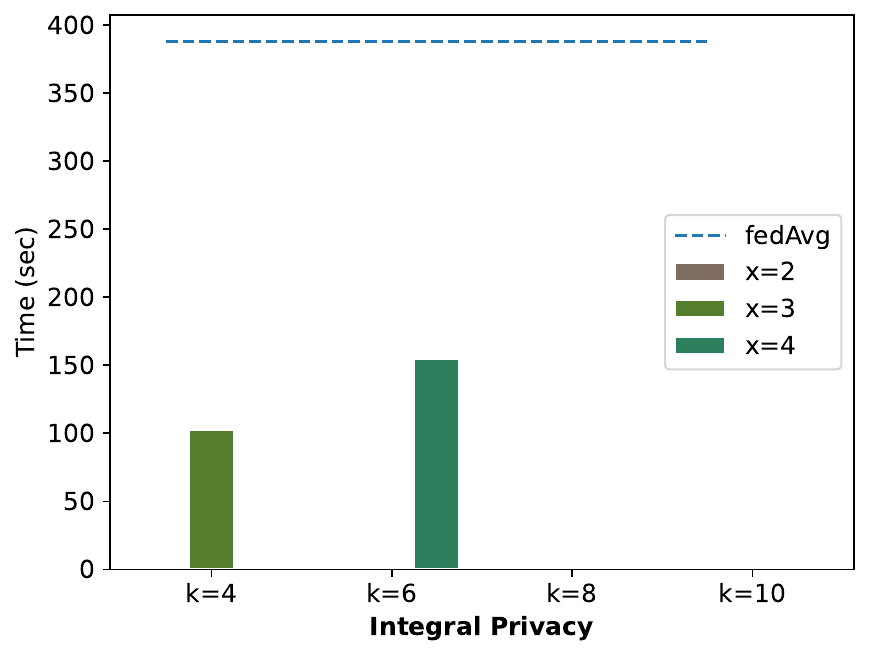}
                \label{MNISTiidunlearn_time}
            }
            \subfigure{
                \includegraphics[width=0.22\textwidth]{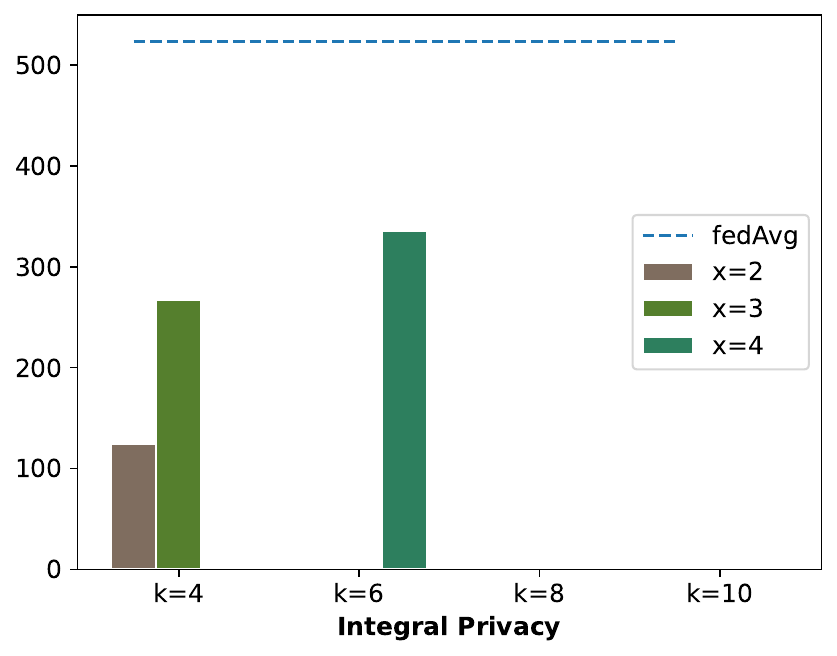}
                \label{FMNISTiidunlearn_time}
            }
            \subfigure{
                \includegraphics[width=0.22\textwidth]{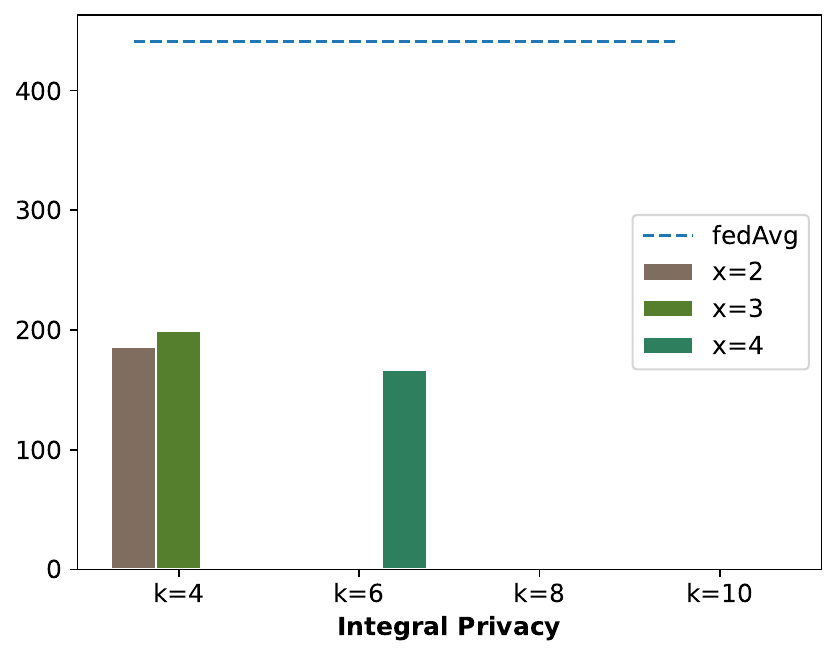}
                \label{CIFAR10iidunlearn_time}
            }
            \subfigure{
                \includegraphics[width=0.22\textwidth]{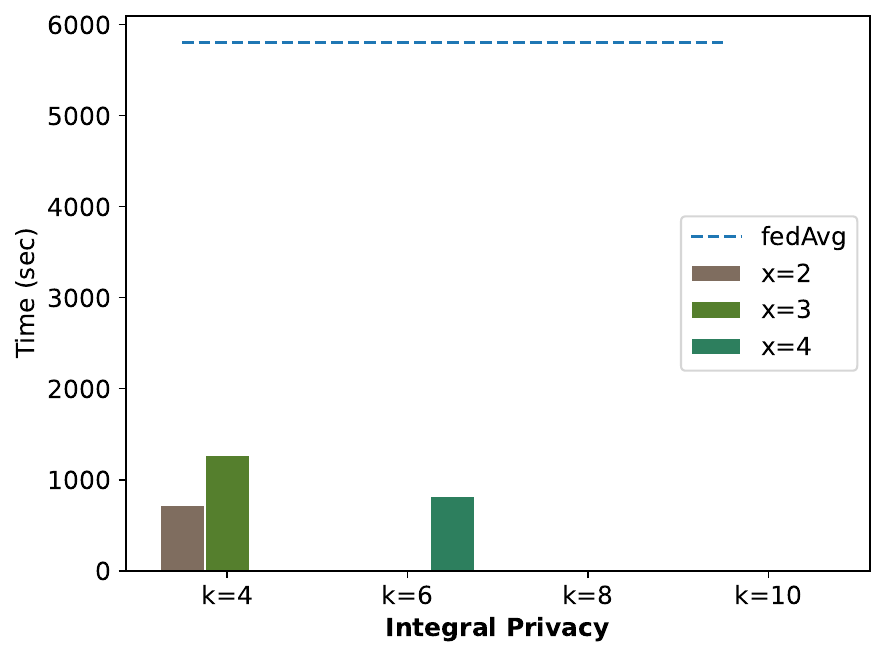}
                \label{celebAiidunlearn_time}
            }
            
            \subfigure{
                \includegraphics[width=0.22\textwidth]{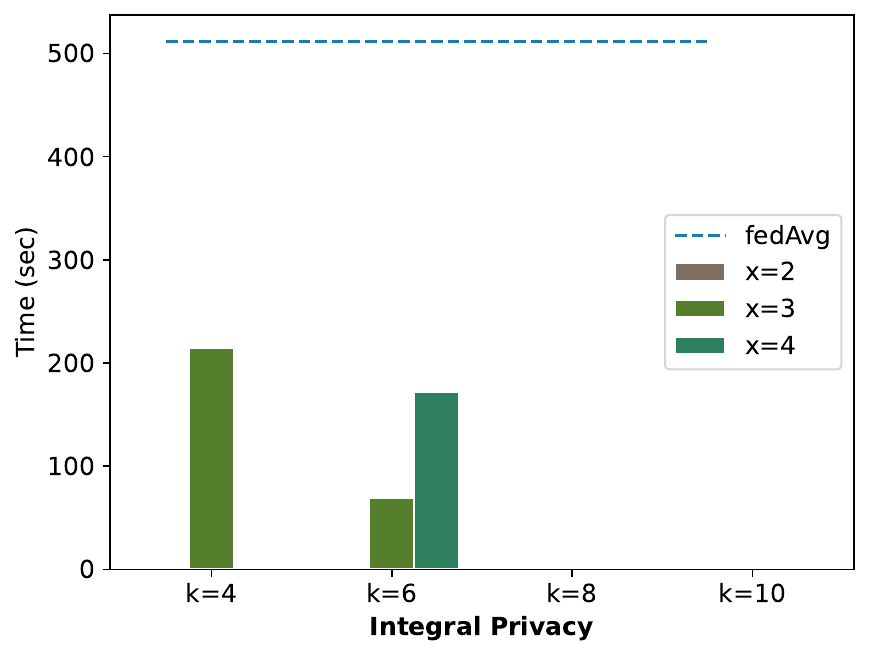}
                \label{MNISTnoniidunlearn_time}
            }
            \subfigure{
                \includegraphics[width=0.22\textwidth]{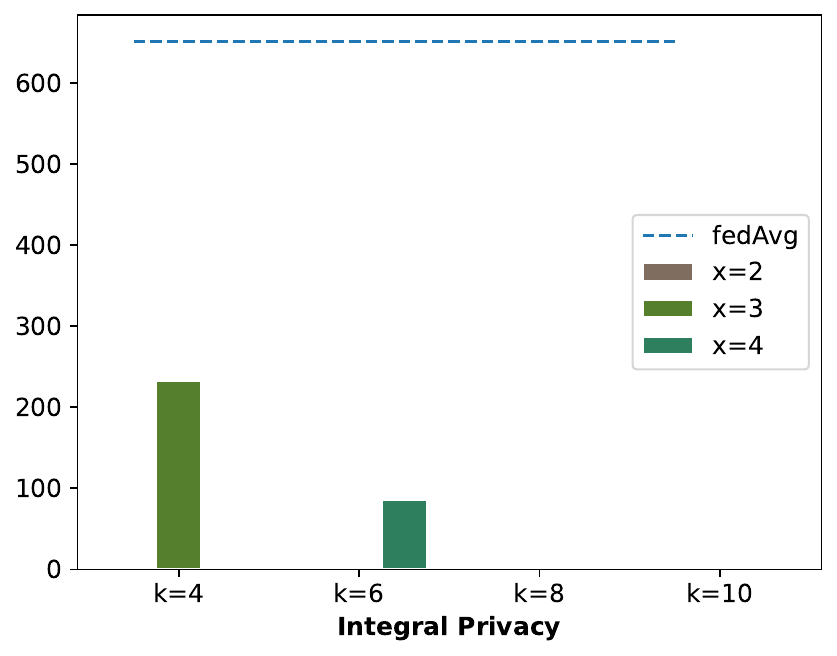}
                \label{FMNISTnoniidunlearn_time}
            }
            \subfigure{
                \includegraphics[width=0.22\textwidth]{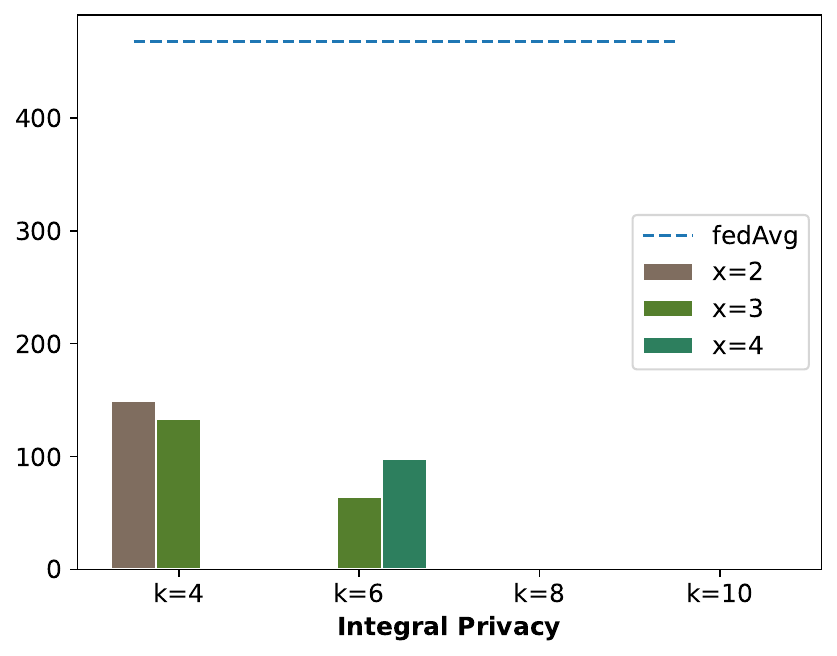}
                \label{CIFAR10noniidunlearn_time}
            }
            \subfigure{
                \includegraphics[width=0.22\textwidth]{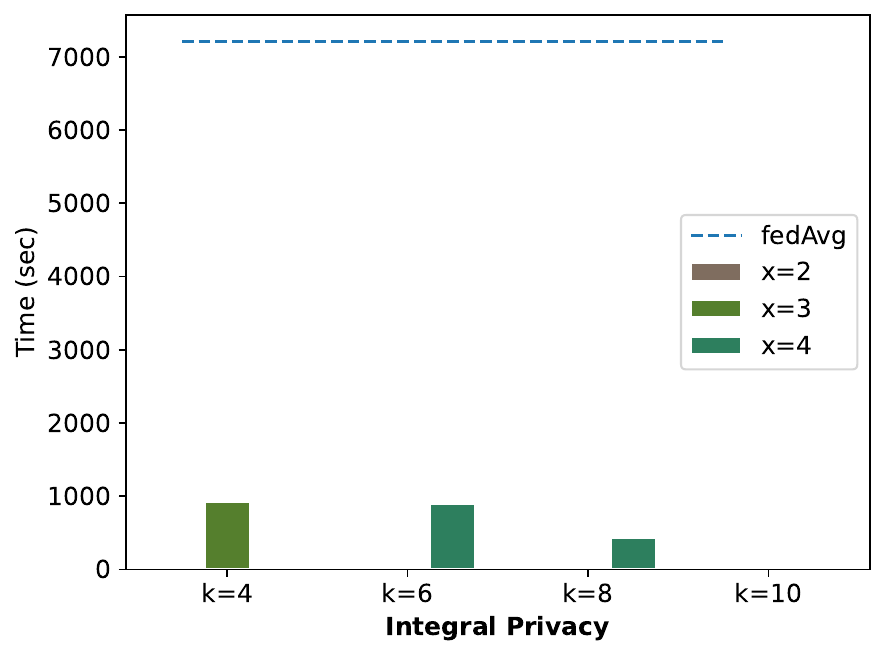}
                \label{celebAnoniidunlearn_time}
            }
        }
\end{figure}

\begin{figure}[ht]
    \floatconts
        {Comparison of add disk space} 
        {\caption{The comparison of Disk Space (y-axis) for $k$-IPfedAvg, and fedEraser for: (a) MNIST-iid (b) FashionMNIST-iid (c) CIFAR10-iid (d) CelebA-iid (e) MNIST-noniid (f) FashionMNIST-noniid (g) CIFAR10-noniid (h) CelebA-noniid.}}
        {
            \subfigure{
                \includegraphics[width=0.22\textwidth]{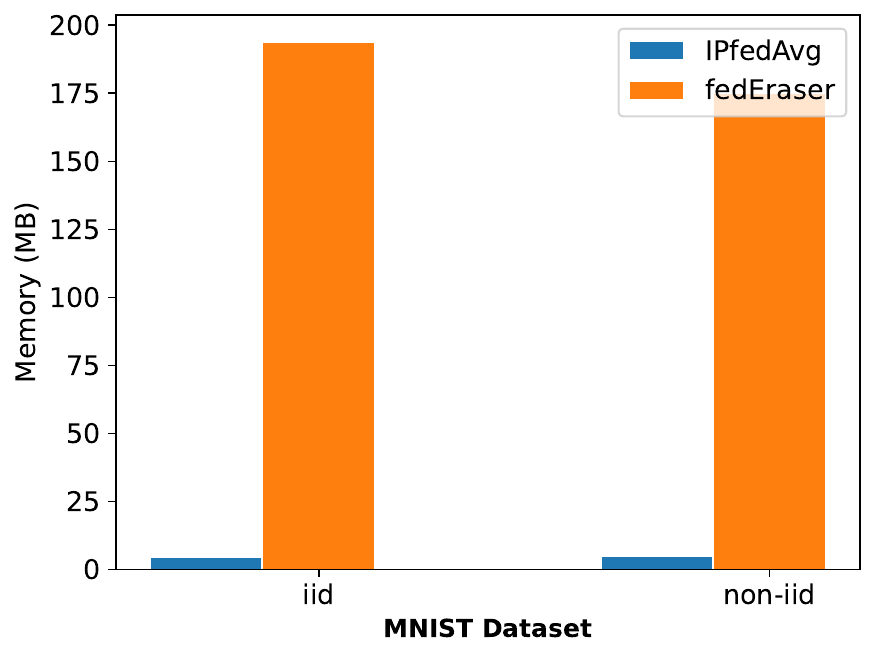}
                \label{MNISTiid_disk_space-LeNet}
            }
            \subfigure{
                \includegraphics[width=0.22\textwidth]{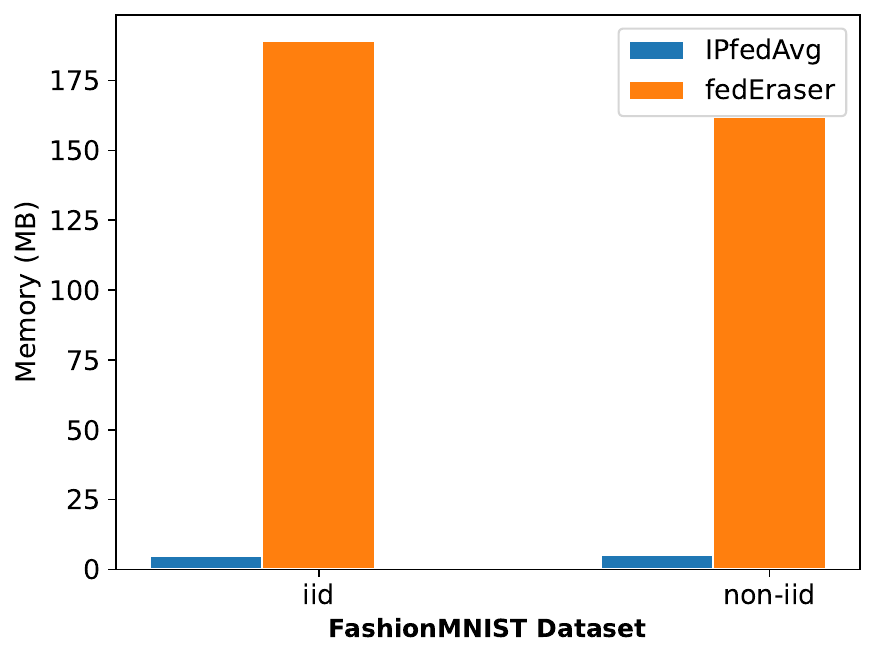}
                \label{FMNISTiid_disk_space-LeNet}
            }
            \subfigure{
                \includegraphics[width=0.22\textwidth]{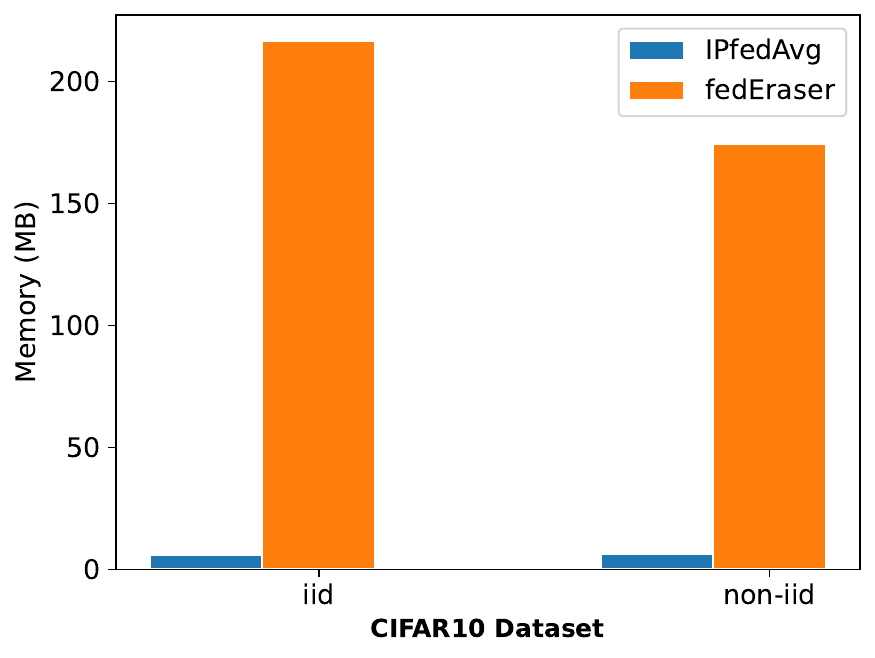}
                \label{CIFAR10iid_disk_space-LeNet}
            }
            \subfigure{
                \includegraphics[width=0.22\textwidth]{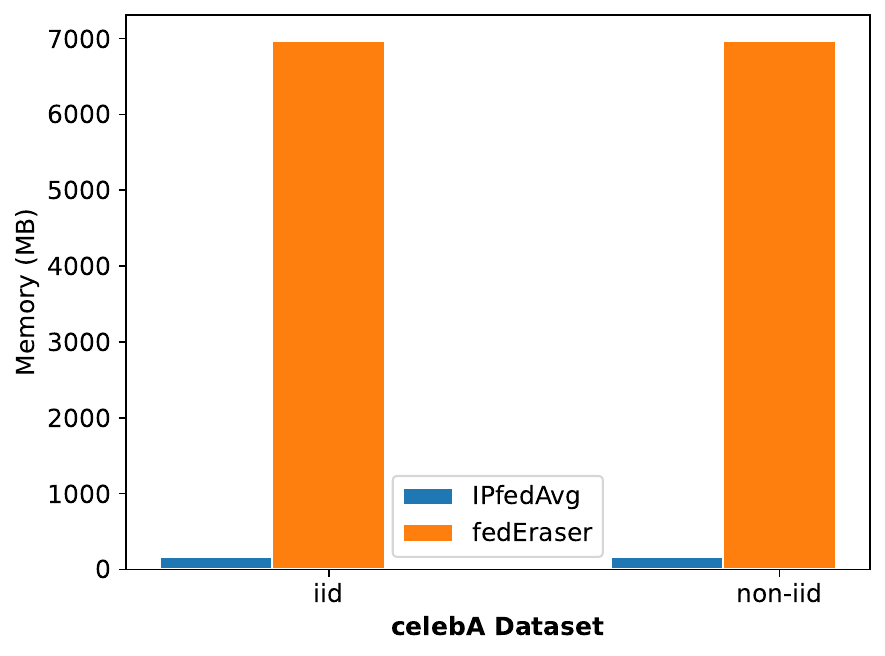}
                \label{CelebAiid_disk_space-LeNet}
            }
            
            \subfigure{
                \includegraphics[width=0.22\textwidth]{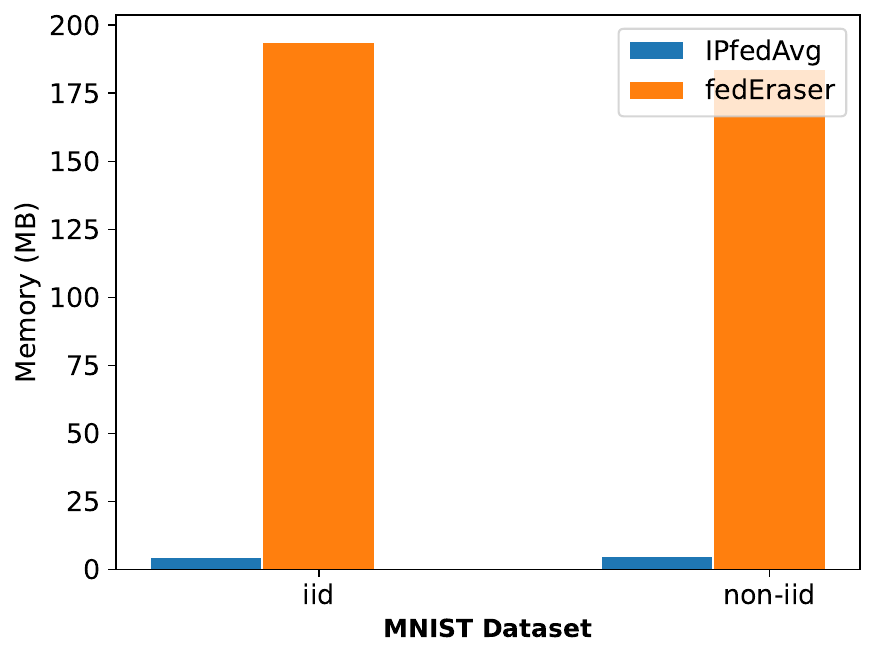}
                \label{MNISTnoniid_disk_space-LeNet}
            }
            \subfigure{
                \includegraphics[width=0.22\textwidth]{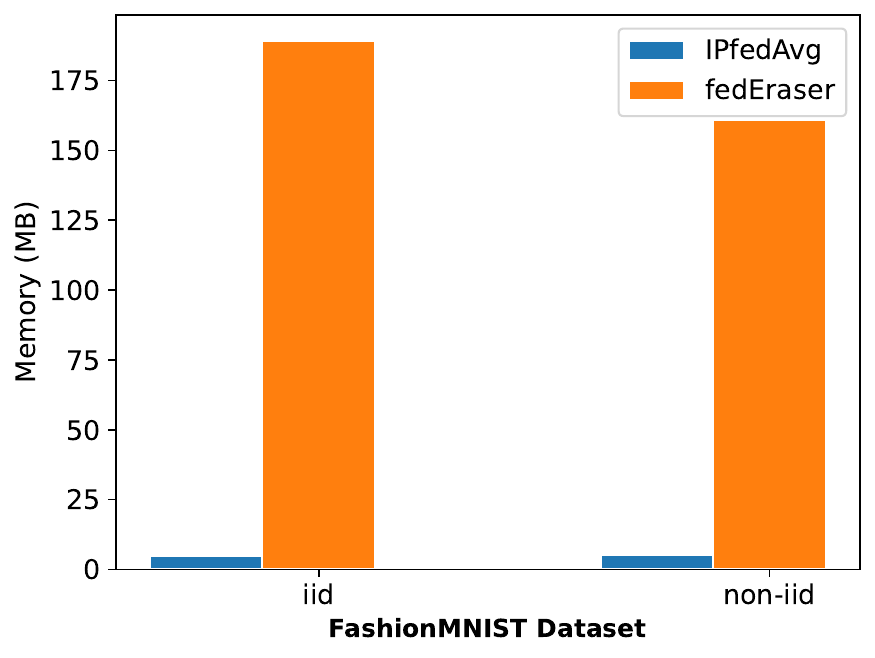}
                \label{FMNISTnoniid_disk_space-LeNet}
            }
            \subfigure{
                \includegraphics[width=0.22\textwidth]{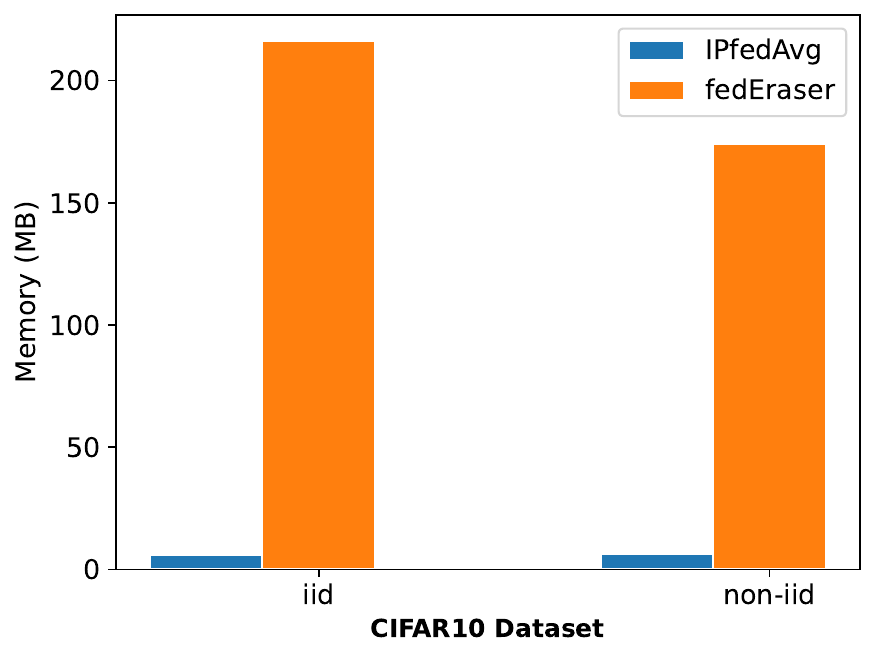}
                \label{CIFAR10noniid_disk_space-LeNet}
            }
            \subfigure{
                \includegraphics[width=0.22\textwidth]{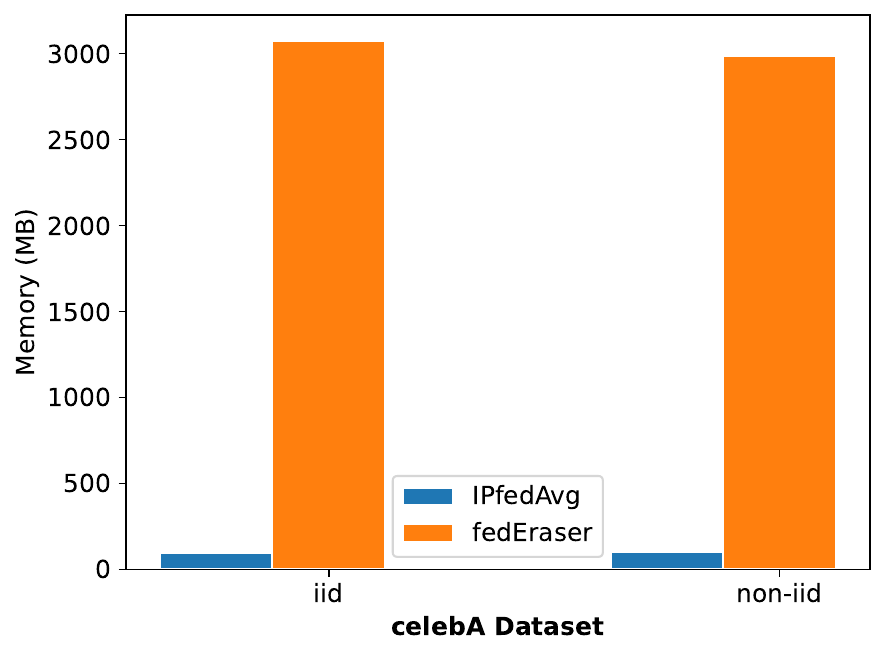}
                \label{celebAnoniid_disk_space-LeNet}
            }
        }
\end{figure}
\end{document}